\documentclass[trackchanges, twocolumn]{aastex701}

\newcommand{\mearth}{M$_{\oplus}$~}

\newcommand{\mjup}{~M$_{J}$}
\newcommand{\ppv}{MgSiO$_3$\,}
\newcommand{\olv}{Mg$_2$SiO$_4$\,}
\newcommand{\apple}{\texttt{APPLE} }
\newcommand{\orchard}{\texttt{ORCHARD} }
\newcommand{\mesa}{\texttt{MESA} }
\newcommand{\mespa}{\texttt{MESPA} }

\newcommand{\dd}{\mathrm{d}}
\newcommand{\pp}{\partial}
\newcommand{\kbbar}{k_{\rm B}/\mathrm{amu}}

\newcommand{\Tint}{T_{\rm int}}
\newcommand{\Teff}{T_{\rm eff}}
\newcommand{\Teq}{T_{\rm eq}}
\newcommand{\Tten}{T_{\rm 10}}

\newcommand{\txt}[1]{\texttt{#1}}

\hypersetup{
  colorlinks=true,
  linkcolor=blue,
  citecolor=blue,
  urlcolor=blue
}

\defcitealias{Sur2024a}{Paper~I}
\defcitealias{Tejada2026a}{TA26}
\defcitealias{Zhang2022}{Z22}
\newcommand{\papI}{\citetalias{Sur2024a}}
\newcommand{\snI}{\citetalias{Tejada2026a}}
\newcommand{\seI}{\citetalias{Zhang2022}}

\usepackage{amsmath}
\usepackage{mathtools}
\usepackage{tikz}
\usetikzlibrary{shapes.geometric, arrows.meta, positioning, calc}
\usepackage[ruled,vlined,linesnumbered]{algorithm2e}
\usepackage{listings}
\begin{document}

\title{\txt{ORCHARD}: A General Planetary Evolution Code}

\correspondingauthor{Roberto Tejada Arevalo}
\email{arevalo@princeton.edu}

\author[0000-0001-6708-3427]{Roberto Tejada Arevalo}
\affiliation{Department of Astrophysical Sciences, Princeton University, 4 Ivy Lane,
Princeton, NJ 08544, USA}
\email[]{arevalo@princeton.edu} 
\author[0000-0002-3099-5024]{Adam Burrows}
\affiliation{Department of Astrophysical Sciences, Princeton University, 4 Ivy Lane,
Princeton, NJ 08544, USA}
\email{burrows@astro.princeton.edu} 
\author[0000-0001-6635-5080]{Ankan Sur}
\affiliation{Department of Earth, Planetary, and Space Sciences, University of California, Los Angeles, 595 Charles E Young Dr E, LA, CA 90095}
\affiliation{Department of Astrophysical Sciences, Princeton University, 4 Ivy Lane,
Princeton, NJ 08544, USA}
\email{ankansur@epss.ucla.edu}
\author[orcid=0000-0001-8283-3425]{Yubo Su}
\affiliation{Canadian Institute for Theoretical Astrophysics, University of Toronto, 60 St George Street, Toronto, M5S 3H8 Ontario, Canada}
\affiliation{Department of Astrophysical Sciences, Princeton University, 4 Ivy Lane,
Princeton, NJ 08544, USA}
\email{yubo.su@utoronto.ca}

\begin{abstract}

We present \texttt{ORCHARD}, a publicly available planetary evolution code based on the gas giant evolution code, \texttt{APPLE}, capable of modeling the evolution and structures of terrestrial, super-Earth, sub-Neptune, Neptune, and gas giant planets and exoplanets from 0.5 M$_\oplus$ to 10 M$_J$. It supports not only the inhomogeneous and non-adiabatic evolution of gas giants and sub-Neptunes, but also the solidification of the mantles and cores of terrestrial planets, sub-Neptunes, and super-Earths. \texttt{ORCHARD} incorporates a state-of-the-art hydrogen-helium equation of state, ``metal" equations of state (water, ice mixtures, enstatite/perovskite, olivine/forsterite, iron), and atmospheric boundary conditions ranging from detailed non-gray radiative transfer models for Solar System giants to irradiated sub-Neptune atmospheres and bare rocky surfaces. \txt{ORCHARD} also supports static calculations with state-of-the-art equations of state. The purpose of \texttt{ORCHARD} is to provide the scientific community with a flexible, unified tool for modeling planetary structures and evolution across the entire mass continuum of general astrophysical and planetary interest.

\end{abstract}


\section{Introduction} 

Nearly sixty years ago, building on earlier equation-of-state (EOS) work from \cite{WignerHuntington1935, DeMarcus1958}, \cite{Hubbard1968} developed the first fully convective models of Jupiter's interior to explain the then-recent thermal emission of Jupiter \citep{Opik1962, Low1966}. This work was expanded to include Saturn \citep{Hubbard1969}, and later extended to the evolution of Uranus and Neptune \citep{Hubbard1978, HubbardMacFarlane1980}. After the discovery of the first exoplanet, 51 Pegasi b \citep{Mayor1995}, \cite{Burrows1995, Saumon1996, Guillot1996}, and \cite{Burrows1997} built upon this early work to create the first gas giant exoplanet and brown dwarf evolutionary models.

Over the past decade, \textit{Juno} \citep{Bolton2017a} and \textit{Cassini} \citep{Matson2003} have revealed that both Jupiter and Saturn likely harbor extended, or ``fuzzy,'' cores \citep{Fuller2014a, Fuller2014b, Wahl2017, Debras2019, Nettelmann2021, Mankovich2021, Militzer2022, Militzer2023, Militzer2024}. Taken together, these observational advances underscore the need for evolutionary models that are self-consistent across the planetary mass range of interest. Beyond the Solar System, \textit{JWST} has ushered in a new era of atmospheric characterization for exoplanets and brown dwarfs, delivering transit and emission spectra that have begun to constrain atmospheric abundances and thermal structures across a wide range of planetary masses \citep[e.g.,][]{Kempton2023a, Wogan2024, Ahrer2025, Barat2025, Madhusudhan2025}. These observations have also motivated new questions about interior composition. For instance, the muted spectral features of sub-Neptunes such as GJ~1214~b \citep{Kempton2023a} and the debated atmospheric detections of K2~18~b \citep{Benneke2019, Wogan2024, Madhusudhan2025} highlight the need for self-consistent interior/atmosphere models that connect interior composition to observable spectra. Near-future missions, including \textit{Ariel} \citep{Tinetti2022, Helled2022a} and the \textit{Habitable Worlds Observatory} \citep{Burns2025}, will extend these capabilities to smaller planets on wider orbits, further expanding the parameter space which interior models must cover. 

The methodological tradition of planetary interior evolution codes, such as \texttt{CEPAM} \citep{Guillot1995_CEPAM}, \apple \citep[][referred to as ``Paper I'' hereafter when referenced individually]{Sur2024a}, and now \mespa \citep[\mesa for planets;][]{Helled2025b}, has heritage in stellar evolution codes \citep{Kippenhahn1990, Kippenhahn2012}, such as \texttt{GARSTEC} \citep{Weiss2008} and \texttt{MESA} \citep{Paxton2011, Paxton2013, Paxton2018}. These evolution codes, however, are complex to develop. Hence, so much of the work on planetary interior evolution has relied on models that assume adiabatic and homogeneous interiors to explore mass-radius relations \citep{Burrows1997,Burrows2001,Fortney2007,Lopez2012a, Lopez2014} and mass-metallicity relations \citep{Thorngren2016, Chachan2025}, and some of these adiabatic models, such as \texttt{GASTLI} \citep{Acuna2024}, are publicly available. Although this latter class of models has been widely applied, their assumptions preclude them from modeling the thermal and compositional evolution of planet interiors via thermal-diffusive processes in an inhomogeneous context, making them unable to properly model the interior structure of the Solar System planets and, perhaps by extension, exoplanets.

To adequately model the non-adiabatic and inhomogeneous evolution of planetary interiors, one must account for energy, thermal, and compositional transport. Planetary interior evolution codes that incorporate the Henyey relaxation method \citep{Henyey1964}, a technique inherited from stellar evolution codes, include \texttt{CEPAM} \citep{Guillot1995_CEPAM}, \texttt{MESPA} \citep{Helled2025b} and \texttt{APPLE} (\papI) for gas giants, and \texttt{CMAPPER}\footnote{The Henyey-evolution code \texttt{CMAPPER} is publicly available here: \url{https://github.com/zhangjis/CMAPPER_rock}.} for super-Earth interior evolution \citep[][``Z22'' hereafter]{Zhang2022}. Such codes have been harnessed to model the likely inhomogeneous evolution of Jupiter and Saturn \citep{Vazan2016, Vazan2018, Muller2020b, Howard2024, Tejada2025, Knierim2025b, Sur2025a, Su2026}, Uranus and Neptune \citep{VazanHelled2020, Tejada2025b, Howard2025}, and gas giant exoplanets \citep{Muller2021, Knierim2024, Knierim2025a, Sur2026a}. However, these existing codes are individually limited in mass range, challenging to use, or are not public.

This paper presents \texttt{ORCHARD}\footnote{\orchard will be available at this GitHub repository upon the acceptance of this manuscript: \url{https://github.com/robtejada/orchard-public}. A frozen version of \txt{ORCHARD} is available at this Zenodo repository: \url{https://zenodo.org/records/22035834}.}, an extended and public version of \texttt{APPLE}, written entirely in \texttt{Python 3}. \orchard covers the terrestrial/rocky mass ranges from 0.5 \mearth up to 10 M$_{J}$ gas giants. \orchard can model the interior thermal evolution of rocky terrestrial planets, super-Earths, sub-Neptunes, and Neptunes, as well as the thermal and compositional evolution of gas giants. Like \texttt{APPLE}, \orchard is built with the flexibility to choose among the most current H-He EOSes, metal ($Z$) EOSes, H-He-Z mixtures at arbitrary $Z$ fractions, choices of atmospheric boundary conditions, and transport quantities, such as Rosseland mean opacities and conductivities. \orchard can model extended compositional gradients in gas giants as well as sub-Neptunes, including convective mixing and erosion of these gradients over time.

As with other planetary evolution code papers \citep[e.g.,][]{Acuna2024, Sur2024a, Helled2025b}, this particular paper does not set out to provide novel scientific results, but instead to present and describe \texttt{ORCHARD}. This paper is organized as follows: Section \ref{sec:methods} describes in detail the structural and evolution equations, energy transport methods, available EOSes, and atmospheric boundary conditions. Section~\ref{sec:examples} provides example calculations for a wide variety of planet masses, and Section~\ref{sec:improvements} describes future upgrades and scientific applications. We provide concluding remarks in Section~\ref{sec:conclusion}. Details on the code algorithm and structure, as well as a flowchart, can be found in the Appendices.
 
\section{Methods}\label{sec:methods}

\subsection{Structure \& Evolution Equations}\label{subsec:structure}

Following the similar methodology of \papI\, and other planetary and stellar evolution codes \citep{Eggleton1971, Demarque2008, Vazan2012, Vazan2013, Paxton2011, Paxton2013, Paxton2018}, \orchard solves the equations of stellar and planetary structure assuming spherical symmetry:

\begin{align}
\frac{dP}{d M_r} &= -\frac{G M_r}{4\pi r^4} + \frac{\omega(t)^2}{6\pi r}
\label{eq:1}\\
\frac{dr}{d M_r} &= \frac{1}{4\pi r^2 \rho}
\label{eq:2}\\
\frac{\partial L}{\partial M_r} &=  -\frac{dU}{dt} - P\frac{d(\frac{1}{\rho})}{dt} + \epsilon_{\rm rad} + \epsilon_L
\label{eq:3}\\
&= -T\frac{dS}{dt} - \sum_i \left(\frac{\partial U}{\partial X_i}\right)_{s,\rho}\frac{dX_i}{dt}
\label{eq:4}\\
        &\qquad\quad
+ \epsilon_{\rm rad} + \epsilon_L
\label{eq:5}\\
N_A\frac{dX_i}{dt} &= -\frac{\partial }{\partial M_r}\left(4\pi r^2 \mathcal{F}_i\right)\, ,
\label{eq:6}
\end{align}
where $M_r$ is the independent Lagrangian coordinate denoting the mass enclosed in a sphere of radius $r$, $P$ is the pressure, $L = 4\pi r^2 \cal{F}$ is the luminosity, $\cal{F}$ is the energy flux, $S$ is the specific entropy, $\rho$ is the mass density, $T$ is the temperature, $U$ is the specific internal energy, $N_A$ is Avogardo's number, $X_i$ is the mass fraction of species $i$, $\mathcal{F}_i$ is the compositional flux (see Section~\ref{subsec:comp_transport}), and $\omega(t)$ is the solid-body angular frequency. 

We evolve $\omega(t)$ to conserve angular momentum. From
the user, a rotation period $P_{\rm rot}$ and a dimensionless
moment-of-inertia coefficient $C_{\rm MoI}$ are used to define a
fixed reference moment of inertia
$I_{\rm ref} \equiv C_{\rm MoI}\, M\, R_p^2$ and a fixed reference
angular frequency $\omega_{\rm ref} = 2\pi/P_{\rm rot}$, where $M$ is
the planet's total mass and $R_p$ is a reference radius
(e.g., the observationally constrained equatorial radius).
Together, these set a fixed reference angular momentum
$L_\omega \equiv I_{\rm ref}\,\omega_{\rm ref}$, which is held
constant throughout the evolution. At every timestep, we recompute
the structural moment of inertia $I(t)$ from the current density
profile. As the planet cools, $\rho(r)$ evolves, changing $I(t)$, and $\omega(t)$ adjusts, $\omega(t) = \omega_{\rm ref}\,I_{\rm ref}/I(t)$, to maintain $L_\omega$. We have embedded into \texttt{ORCHARD} the Theory of Figures to fourth order
\citep[ToF4;][]{Zharkov1975, Nettelmann2017} to calculate the moment of inertia $I(t)$ at each timestep.\footnote{We also use ToF4 to calculate the gravity moments $J_2$ and $J_4$ at each timestep. In the future, we plan to expand our formalism to include the Theory of Figures to seventh order \citep{Nettelmann2021}, which is required to more accurately calculate higher-order moments.} This feature is provided to enable comparisons between models and the moments measured by satellite probes such as \textit{Juno} and \textit{Cassini}, and can be activated by setting \texttt{rotation = True} and \texttt{tof\_calc = True} in the user parameter configuration file (See Appendix B). 

These coupled equations support not just evolution calculations, but static profiles. Equation~\ref{eq:3} is the standard energy equation, while Equation~\ref{eq:4} accounts for compositional changes with the second term. The term $\epsilon_{\rm rad}$ is the specific radiogenic heating rate from the decay of long-lived radioactive isotopes, and $\epsilon_L$ is the specific latent heat release rate associated with first-order phase transitions such as mantle and core solidification. These luminosity sources are applied only to the mantles and cores. We describe Equations~\ref{eq:4} and \ref{eq:5} in more detail in Section~\ref{subsec:transport}, and Equation~\ref{eq:6} in Section~\ref{subsec:comp_transport}. We refer the reader to \papI\, for a more detailed description of these equations and their application in \apple and by inheritance in \texttt{ORCHARD}. A detailed description of our Henyey relaxation and Newton-Raphson solvers is provided in Appendix A.

\subsection{Energy Transport}\label{subsec:transport}

As described in \papI\, and in \citet[][``TA26'' hereafter]{Tejada2026a}, heat transport in \apple is modeled by the total heat fluxes of radiation, conduction, and convection, 

\begin{equation}\label{eq:total_flux}
    \mathcal{F}_{\rm tot} = \mathcal{F}_{\rm rad} + \mathcal{F}_{\rm cond} + \mathcal{F}_{\rm conv}\, ,
\end{equation}
where $\mathcal{F}_{\rm rad}$ is the diffusive radiative heat flux, $\mathcal{F}_{\rm cond}$ is the conductive heat flux, and $\mathcal{F}_{\rm conv}$ is the convective heat flux. This section briefly describes each component.

\subsubsection{Radiation \& Conduction}

The radiative flux is given by \citep[e.g.,][]{Kippenhahn2012}:
 \begin{equation}
     \mathcal{F}_{\rm rad} = -\frac{4ac}{3}\frac{T^3}{\kappa_r \rho}\frac{\partial T}{\partial r} = -\lambda_r \frac{\partial T}{\partial r}\, , 
     \label{eq:rad_flux}
 \end{equation}
 where $\kappa_r$ is the Rosseland mean opacity, $a$ is the radiation constant, $c$ is the speed of light, $T$ is the local temperature, $\rho$ is the local mass density, $\partial T/\partial r$ is the local temperature gradient, and $\lambda_r = \frac{4ac}{3}\frac{T^3}{\kappa_r \rho}$ is the associated ``thermal conductivity" due to photon heat transport. 
 
 The heat flux due to conduction is given by:

 \begin{equation}\label{eq:f_cond}
     \mathcal{F}_{\rm cond} = -\lambda_{\rm cd} \frac{\partial T}{\partial r}\, ,
 \end{equation}
where $\lambda_{\rm cd}$ is the thermal conductivity of the conducting material. For H-He mixtures, we include the density- and temperature-dependent conductivities from \cite{French2012} and \cite{Becker2018}. For water, we use the density- and temperature-dependent thermal conductivities calculated by \cite{French2019}, and for rocks (either \ppv or Mg$_2$SiO$_4$), we use the density- and temperature-dependent conductivities of \cite{Stamenkovic2011}. For iron, we set the default standard conductivity to 40 W m$^{-1}$ K$^{-1}$, which is consistent with the estimated thermal conductivity of the Earth's core \citep{Luo2024}. The user can set their desired mantle and core conductivities by using the configuration parameters \texttt{mantle\_thermal\_conductivity} and \texttt{core\_thermal\_conductivity}. These parameters set constant thermal conductivities for the mantle and core, respectively, in the user parameter configuration file (See Appendix B). 

\subsubsection{Convective Energy Transport}

We apply the Mixing-Length Theory (MLT) approach \citep{Bohm1958} to handle convective transport. In traditional MLT, the convective flux is described in terms of the average vertical distance traveled before a convective fluid parcel equilibrates with its surroundings. For the purposes of derivation, we assume a uniform composition. The convective flux, therefore, is assumed to be

\begin{equation}\label{eq:f_conv_mlt}
    \mathcal{F_{\rm conv}} = \rho v_{\rm MLT} C_P\Delta T\, ,
\end{equation}
where $C_P$ is the isobaric heat capacity and $\Delta T$ is the temperature change of the fluid parcel due to its vertical motion. The mixing length velocity $v_{\rm MLT}$ is defined as \citep[see Equation 7.6 of][]{Kippenhahn2012}:

\begin{equation}\label{eq:v_mlt}
    v^2_{\rm MLT} = g\delta(\nabla - \nabla_{\rm ad})\frac{l^2}{8 H_p}\, ,
\end{equation}
where $g$ is the gravitational acceleration, $\delta = (\partial \ln{\rho}/\partial \ln{T})_P$, and $l$ is the so-called ``mixing length.'' The pressure scale height, $H_p$, is defined as $H_p = -dr/d\ln{P}$. Typically, $l = \alpha_{\rm MLT}H_P$, where $\alpha_{\rm MLT}$ is referred to as the ``mixing length parameter.'' The temperature gradient of the structure and the adiabatic temperature gradient are $\nabla = d\ln{T}/d\ln{P}$ and $\nabla_{\rm ad} = (\partial \ln{T}/\partial \ln{P})_{S}$, respectively. The change in temperature $\Delta T$, can be approximated as

\begin{equation}\label{eq:delta_T}
    \Delta T = (\nabla - \nabla_{\rm ad})\frac{l}{2}\frac{T}{H_P}\, .
\end{equation}

If we set $l = H_p$, then placing Equations~\ref{eq:v_mlt} and \ref{eq:delta_T} into Equations~\ref{eq:f_conv_mlt} yields

\begin{equation}\label{eq:fconv_nabla}
    \mathcal{F_{\rm conv}} = \rho C_PT\sqrt{\frac{\delta g H_p}{32}}(\nabla - \nabla_{\rm ad})^{3/2}\, ,
\end{equation}
which is positive if the Schwarzschild condition \citep[$\nabla > \nabla_{\rm ad}$;][]{Schwarzschild1906} is satisfied. Assuming uniform composition, it can be shown that \citep[see Section 6.2 of][for a detailed derivation]{Tejada2024}

\begin{equation}
    \nabla - \nabla_{\rm ad} = -\frac{H_p}{C_p}\frac{dS}{dr}\, ,
\end{equation}
where $dS/dr$ is the structural specific entropy gradient. Hence, Equation~\ref{eq:fconv_nabla} can be expressed as 

\begin{equation}\label{eq:fconv_apple}
    \mathcal{F_{\mathrm{conv,invisc}}} = -\rho C_pT\sqrt{\frac{\delta g H_p}{32}}\bigg(\frac{H_p}{C_p}\frac{dS}{dr}\bigg)^{3/2}\, ,
\end{equation}
which can be expanded to encompass the Ledoux criterion \citep{Ledoux1947} for convection, as shown in Appendix A of \cite{Tejada2024}. Equation~\ref{eq:fconv_apple} is the convective flux formalism used in \papI, implemented in \apple and, therefore, also in \texttt{ORCHARD}. The ``invisc'' subscript in Equation~\ref{eq:fconv_apple} indicates that this convective flux applies in the \textit{inviscid} limit to distinguish it from the \textit{viscous} convection limit (see next Section).

In the presence of compositional gradients, where $X_i$ are the mass fractions of
each species, the Ledoux condition for convection can be expressed as 
\begin{equation}\label{eq:ledoux_condition}
    -\frac{H_P}{C_p}\bigg[\frac{dS}{dr} - \sum_i\bigg(\frac{\partial S}{\partial X_i}\bigg)_{\rho, P}\frac{dX_i}{dr}\bigg] > 0\, ,
\end{equation}

We use Equation~\ref{eq:fconv_apple} throughout \orchard to model convective fluxes across all regions, and expand to the Ledoux version when considering compositional gradients. Note that mantles and cores in \orchard do not experience compositional exchange, but incorporating this
process is one goal for future upgrades
(see Section~\ref{subsec:comp_exchange}).

\subsubsection{Mantle viscous convection}\label{subsec:visc_conv}

In \snI, we updated \apple to model viscous convection following the
methodologies of \cite{Sasaki1986a} and \seI. In the viscous-limited regime,
such as convection of molten magma in the Earth's mantle, the convective velocity of the fluid parcels is linear with $\nabla - \nabla_{\rm ad}$ due to the drag and buoyancy force balance. It can be expressed as 

\begin{equation}\label{eq:v_misc}
    v_{\rm visc} = g\delta(\nabla - \nabla_{\rm ad})\frac{l^3}{16\nu H_p}
\end{equation}
where $\nu$ is a general kinematic viscosity.\footnote{See Appendix C for a derivation of \ref{eq:v_misc}.} The change in temperature in Equation~\ref{eq:delta_T} is a general thermodynamic statement, so combining Equations~\ref{eq:delta_T} and \ref{eq:v_misc} into the general convective MLT flux in Equation~\ref{eq:f_conv_mlt} yields

\begin{equation}
    \mathcal{F}_{\rm conv,\,visc} = \rho\,C_P\,T\,\frac{g\,\delta\,H_P^2}{32\,\nu_{\rm solid}}\,(\nabla-\nabla_{\rm ad})^2\,.
\end{equation}

In terms of the entropy gradients,\footnote{Note that the convection criterion $dS/dr <0$ makes Equation~\ref{eq:fconv_apple} positive, hence both Equations~\ref{eq:f_conv_mlt} and \ref{eq:fconv_visc} are positive.} the flux can then be expressed as 

\begin{equation}\label{eq:fconv_visc}
    \mathcal{F_{\rm conv,visc}} = \rho C_p T\,\frac{\delta\, g\, H_p^2}{32\,\nu_{\rm solid}}\,
    \bigg(\frac{H_p}{C_p}\frac{dS}{dr}\bigg)^2\, ,
\end{equation}
where $\nu_{\rm solid}$ is the kinematic viscosity
of the solid phase, which is calculated from the dynamic viscosity, $\eta_{\rm solid}$, with $\nu_{\rm solid} = \eta_{\rm solid}/\rho$. The dynamic viscosity of solid silicate follows a dislocation-creep law
\citep{Ranalli2001}:
\begin{equation}\label{eq:eta_solid}
  \eta_{\rm solid}
  = \frac{1}{2}\,B^{-1/n}\,\dot\varepsilon^{(1-n)/n}
    \exp\!\left(\frac{E^* + PV^*}{nRT}\right)\, ,
\end{equation}
where $B$, $n$, $E^*$, and $V^*$ are material-dependent parameters and $\dot\varepsilon=10^{-15}$~s$^{-1}$ is the assumed strain rate. The dislocation-creep parameters for each silicate phase are listed in Table~\ref{tab:creep_params}.

\begin{deluxetable}{lcccc}
\tablecaption{Dislocation-creep parameters for Equation~\ref{eq:eta_solid} \citep{Ranalli2001}.\label{tab:creep_params}}
\tablecolumns{5}
\tablehead{
  \colhead{Material} &
  \colhead{$B$ [Pa$^{-n}$\,s$^{-1}$]} &
  \colhead{$n$} &
  \colhead{$E^*$ [erg\,mol$^{-1}$]} &
  \colhead{$V^*$ [cm$^3$\,mol$^{-1}$]}
}
\startdata
Olivine (Mg$_2$SiO$_4$)    & $3.5\times10^{-15}$ & 3.0 & $4.3\times10^{12}$ & 10 \\
Perovskite (MgSiO$_3$)     & $7.4\times10^{-17}$ & 3.5 & $5.0\times10^{12}$ & 10 \\
\enddata
\end{deluxetable}

The effective kinematic viscosity of a partially molten region of the mantle is combined logarithmically as
\begin{equation}\label{eq:nu_blend}
  \log_{10}\nu_{\rm eff}
  = \chi\,\log_{10}\nu_{\rm liq}
  + (1-\chi)\,\log_{10}\nu_{\rm solid}\, ,
\end{equation}
where $\chi$ is the melt fraction profile of the mantle and core. The dynamic viscosity of liquid rock is assumed to be $10^{3}$~poise \citep{Abe1997, Harris2008}. The melt fraction, $\chi$, is defined as a smooth function of pressure and temperature in the silicate mantle and iron-rich core sections of the structure,

\begin{equation}\label{eq:melt_fraction}
    \chi(P,T) = \frac{1}{2}\left[1 + \tanh\!\left(\frac{T - T_{\rm melt}(P)}{\Delta T}\right)\right]\, ,
\end{equation}
where $\Delta T$ is a smoothing width (default $\Delta T = 200\;\mathrm{K}$).\footnote{Not applied to H-He or water envelopes.} The weight satisfies $\chi \to 0$ deep in the solid phase ($T \ll T_{\rm melt}$) and $\chi \to 1$ deep in the liquid phase ($T \gg T_{\rm melt}$). The melt temperature ($T_{\rm melt}$) depends on whether the user chooses \olv or \ppv for their mantle composition by using \texttt{mantle\_comp=mg2sio4} or \texttt{mantle\_comp=mgsio3} in their parameter configuration file (See Appendix B). These EOSes are described in Section~\ref{subsubsec:pure_metal}.

Following \seI, during mantle solidification, each mass zone experiences a transition between the fully liquid (inviscid, $\chi=1$) and fully solid
(viscous, $\chi=0$) convective regimes. As a fluid parcel rises in the partially molten state, its temperature and melt fraction change. Consequently, the change in melt fraction informs the change in the density (not just the temperature change) of the parcel fluid and, hence, the convective flux in these regions. Instead of the thermal expansion coefficients, the convective fluxes at the melt curves depend on the thermal expansion due to the phase change, $\alpha_\chi \equiv -(1/\rho)(\partial \rho/\partial\chi)_P$. The convective fluxes at the melt curve, then, are defined as 
\begin{align}
    \mathcal{F}_{\mathrm{invisc},T} &= -\rho\,\mathcal{L}\sqrt{\frac{\alpha_\chi\,g\,H_p}{32}}
        \bigg(\frac{H_p\,T_{\rm melt}}{\mathcal{L}}\frac{dS}{dr}\bigg)^{3/2}\,,
        \label{eq:fi_melt}\\
    \mathcal{F}_{\mathrm{visc},T} &= -\rho\,\mathcal{L}\,
        \frac{\alpha_\chi\,g\,H_p^2}{32\,\nu_{\rm eff}}
        \bigg(\frac{H_p\,T_{\rm melt}}{\mathcal{L}}\frac{dS}{dr}\bigg)^{2}\,,
        \label{eq:fv_melt}
\end{align}
where $\mathcal{L}$ is the latent heat released along the melt curve.\footnote{The factor of 32 arises due the heat exchanged by the parcel carries the mean-displacement factor of $1/2$ (Equation~\ref{eq:delta_T}).} A more detailed explanation of Equations~\ref{eq:fi_melt} and \ref{eq:fv_melt} is provided in Appendix~\ref{sec:melt_flux_derivation}.  To capture the effects of rising and sinking parcels of fluid at the melt curve, we linearly blend the local and melt-curve fluxes by melt
fraction:
\begin{align}
    \mathcal{F}_{\rm eff, invisc} &= \chi\,\mathcal{F_{\mathrm{conv,invisc}}} + (1-\chi)\, \mathcal{F}_{\mathrm{invisc},T}
        \,,\label{eq:fi_eff}\\
    \mathcal{F}_{\rm eff, visc} &= (1-\chi)\,\mathcal{F_{\mathrm{conv,visc}}} + \chi\,\mathcal{F}_{\mathrm{visc},T}
        \,.\label{eq:fv_eff}
\end{align}

Because the dynamic viscosity spans $\sim 10^{3}$--$10^{13}$~poise over the
range of mantle conditions, we combine Equations~\ref{eq:fi_eff} and \ref{eq:fv_eff} via a geometric mean, yielding the effective melt-zone convective flux
\begin{equation}\label{eq:fconv_melt}
    \mathcal{F}_{\rm conv,melt} =
        \mathcal{F}_{\rm eff, invisc}^{\chi}\mathcal{F}_{\rm eff, visc}^{1-\chi}
        \,.
\end{equation}

We apply Equation~\ref{eq:fconv_melt} throughout the evolution of the mantles of sub-Neptunes and super-Earths. This reduces to Equation~\ref{eq:fconv_apple} in fully liquid ($\chi=1$) regions, and reduces to Equation~\ref{eq:fconv_visc} in fully solid ($\chi = 0$) regions. We emphasize here that these combinations are applied locally within the mantle, not to the iron-rich cores. Iron core convection is driven by Equation~\ref{eq:fconv_apple} in the liquid state, and we halt such convection when iron solidifies, thereby making a solid and conductive iron core.

\subsubsection{Latent \& Radiogenic Heat}
Latent and radiogenic heat are source terms to Equation~\ref{eq:4}. The latent heat of melting and solidification is evaluated only in the silicate mantle (Equations~\ref{eq:tmelt_mgsio3} and \ref{eq:tmelt_mg2sio4}) and in the iron core (Equation~\ref{eq:tmelt_fe}).\footnote{While there is a latent heat release associated with helium rain in the H-He-Z envelopes of gas giants \citep[e.g.,][]{Markham2024a}, this latent heat is not yet included since estimates are not yet available.} Equation~\ref{eq:4} can then be expressed as the entropy update:

\begin{equation}\label{eq:entropy_master}
\begin{split}
  \frac{d S}{d t} = & -\frac{4\pi r^2}{T}\frac{\partial\mathcal{F_{\rm tot}} }{\partial m}
  + \overbrace{
      \frac{\mathcal{L}}{T}\frac{d\chi}{dt} + \frac{H(t)}{T}
    }^{\mathclap{\text{Latent \& radiogenic heat}}} \\[2ex]
  & - \overbrace{
      \frac{1}{T}\left[ \sum_i\bigg(\frac{\pp U}{\pp X_i}\bigg)_{S,\rho}\frac{d X_i}{d t} \right]
    }^{\mathclap{\text{Chemical potential terms}}}\, ,
\end{split}
\end{equation}
where $H(t)$ is the specific radiogenic heat. The term $d\chi/dt$ is the rate of change of the melt fraction. Following \seI, radioactive decay in the rocky mantle contributes a specific heating rate
(erg\;s$^{-1}$\,g$^{-1}$):
\begin{equation}
\begin{gathered}
H(t) = \sum_i w_i\, q_{0,i}\, \exp\!\left[\ln 2\,\frac{(t_\oplus - t)}{\tau_i}\right], \\
t_\oplus = 4.567\ \mathrm{Gyr},
\end{gathered}
\label{eq:Hspec}
\end{equation}
from the decay chains of $^{238}$U, $^{235}$U, $^{232}$Th, and $^{40}$K, scaled with abundances. The $w_i$ are dimensionless scale factors for each radiogenic isotope relative to the Earth-mantle abundances of \cite{McDonough1995}. The $q_{0,i}$ are the present-day heat production rates corresponding to each species per unit of Earth's mantle mass, $\tau_i$ is the half-life of such species, and $t$ is the current model age. These values are tabulated in Table~\ref{tab:isotopes}. The latent heat and radiogenic heat effects can be disabled by setting \texttt{latent\_heat\_effects = False} and \texttt{radioactive = False} in a user's configuration file.  

\begin{deluxetable}{lcc}
\tablecaption{Parameters of Radiogenic Elements \citep{McDonough1995}.\label{tab:isotopes}}
\tablecolumns{3}
\tablehead{
  \colhead{Element} &
  \colhead{$q_0$ [erg\,s$^{-1}$\,g$^{-1}$]} &
  \colhead{$\tau$ [Gyr]}
}
\startdata
$^{40}$K    & $8.69\times10^{-9}$  & 1.25  \\
$^{232}$Th  & $2.24\times10^{-8}$  & 14    \\
$^{235}$U   & $8.48\times10^{-10}$ & 0.704 \\
$^{238}$U   & $1.97\times10^{-8}$  & 4.47  \\
\enddata
\tablecomments{Radiogenic heating isotopes used in Equation~\ref{eq:Hspec}. These isotopes are assumed to be evenly spread throughout the mantle.}
\end{deluxetable}

 \subsection{Compositional Transport}\label{subsec:comp_transport}

 Convective regions homogenize their local compositions on timescales shorter than evolutionary timescales. As in \papI, we solve the diffusion equation, 

 \begin{equation}\label{eq:conv_mixing}
    \frac{d X_i}{d t} = \frac{\partial}{\partial M_r}\bigg(4\pi r^2 \rho \mathcal{D}\frac{\partial X_i}{\partial r}\bigg)\, ,
\end{equation}
where $M_r$ is the mass shell at radius $r$ and $\mathcal{D}$ is the convective diffusion coefficient defined as $\frac{1}{3}\ v_{\rm MLT} l$. 

To model the effects of helium rain expected in gas giant planets and water and silicate rain in sub-Neptune exoplanets (\snI; future \orchard upgrade), an advection term is added to Equation~\ref{eq:conv_mixing} to model phase separation of either helium \citep{Stevenson1975}, water \citep{Bergermann2024, Gupta2025}, or silicates \citep{StixrudeGilmore2025, Rogers2025}. Diffusion-advection methods for helium rain were originally developed in \cite{Sur2024a} and were first applied to evolutionary models of Jupiter and Saturn in \cite{Tejada2025} and \cite{Sur2025a}. Equation 49 of \papI\, (Scheme B) was expanded into the $Z$ component to model silicate rain in \snI. We further generalize Scheme B of \papI\, here, where the $X_i$ component can be phase-separated independently and driven to its equilibrium abundances informed by the respective miscibility curve, $X_{i,\rm{low}}$ and $X_{i,\rm{high}}$:

\begin{equation}\label{eq:misc}
    \begin{aligned}
        \frac{d X_i}{d t} 
        &= \frac{\partial}{\partial M_r}\bigg\{
        4\pi r^2 \rho \bigg[\mathcal{D}\frac{\partial X_i}{\partial r}\\
        &\qquad\quad
        + v_{\rm rain}\max(0,X_i - X_{i,\rm low}) \max(0,X_{i,\rm high} - X_i)\bigg]\bigg\}\, .
    \end{aligned}
\end{equation}
Here, the rain sedimentation velocity, $v_{\rm rain}$, is defined as

\begin{equation}\label{eq:v_sed}
    v_{\rm rain} = \frac{\mathcal{D}}{\alpha_{\rm rain} H_p}\, ,
\end{equation}
where $\alpha_{\rm rain}$ is the miscibility/phase separation rain scale height factor. Lower values yield higher sedimentation velocities (e.g., $\sim$1 cm s$^{-1}$) and, thus, thinner rain regions and greater depletion. Since advection is guided by the phase equilibrium or immiscibility abundances, advection in Equation~\ref{eq:misc} is only active when local abundance, $X_{i,\rm high} > X_i > X_{i,\rm low}$, since abundances outside of these bounds are miscible. We demonstrate its effects in Section~\ref{sec:examples}.  The low and high equilibrium abundances ($X_{i,\rm high}$ and $X_{i,\rm low}$) are given by the miscibility curves of hydrogen-helium \citep{Lorenzen2009, Lorenzen2011}, hydrogen-water \citep{Gupta2025}, or hydrogen-silicates \citep{StixrudeGilmore2025, Rogers2025}. We show these curves for helium in Figure 2 of \cite{Tejada2024} and for silicates in Figure 4 of \snI. The user can choose between the miscibility curves of \cite{Lorenzen2009, Lorenzen2011} and those of \cite{Schottler2018} by setting \texttt{misc\_curve = l} or \texttt{misc\_curve = s}, respectively. 

As described in \snI, the local abundance, $X_i$, is driven to either $X_{i,\rm{low}}$ or $X_{i,\rm{high}}$, depending on their location in the interior structure, and the rate at which we rainout the local $Z$ is proportional to the difference between $X_i$ and its equilibrium value (e.g., $X_i-X_{i,\rm{low}}(P, T)$). As such, depletion is not instantaneous and depends on the distance from the local coexistence curves and on feedback from surrounding convective regions.

\subsection{Equations of State}\label{subsec:EOS}

The wide range of planetary masses modeled by \orchard requires various equations of state for use in the envelopes, mantles, and cores. The mantles and cores of terrestrial, super-Earth, and sub-Neptune planets are treated similarly as ``compact'' cores of gas and ice giant planets.\footnote{We provide a central and comprehensive EOS module used by \orchard here for public use: \url{https://github.com/robtejada/eos/tree/eos_orchard}. The first version of this EOS module was published by \cite{Tejada2024}. This module contains various H-He and $Z$ equations of state for general planetary interior modeling.} We focus here on the description and usage of these EOSes for planetary interior characterization, ranging from H-He mixtures and silicate mantles to iron-rich cores. A comprehensive summary description is found in Table~\ref{tab:eos_summary}.

\subsubsection{H-He mixture EOSes with non-ideal effects}\label{subsec:xy_mix}

The default H-He EOS used in \orchard is that of \citet[][CD21]{Chabrier2021}, which is derived from the ab initio EOSes of \cite{Militzer2013} combined with \citet[][CMS19]{Chabrier2019}. The H-He EOS of \cite{Militzer2013} was calculated with the non-ideal entropy and density interactions, but is limited in scope. On the other hand, the comprehensive CMS19 EOS spans the entire thermodynamic space and can be calculated at any helium fraction. We also provide the CMS19 EOS that includes the non-ideal density, entropy, and internal energy corrections of \citet[][CMS19+HG23]{Howard2023a} and uses the H-He EOS of \citet[][SCvH95]{Saumon1995} at densities lower than 0.05 g cm$^{-3}$. Users can use either the CD21, CMS19 (with non-ideal corrections incorporated), or the SCvH95 EOSes by selecting \texttt{hhe\_eos = cms}, \texttt{hhe\_eos = cd}, or \texttt{hhe\_eos = scvh}, respectively, in their configuration files. The option to use the SCvH95 is primarily for comparison purposes.

To summarize, we obtain the density, specific (i.e., per unit mass) entropy, and specific internal energy of H-He mixtures at any helium mass fraction ($Y$) and employ the volume-addition law. The specific volume ($V_{xy}$) of the H-He mixture is

\begin{equation}\label{eq:vxy}
    V_{xy} = \frac{1-Y'}{\rho_{\rm H}(P,T)} + \frac{Y'}{\rho_{\rm He}(P,T)} + \Delta V_{\rm mix}(P,T,Y')\, ,
\end{equation}
where $Y' = Y/(X + Y)$, or equivalently, in the presence of metals ($Z$), $Y' = Y/(1 -Z)$. Here $\Delta V_{\rm mix}$ is the excess or deficit of volume due to the non-ideal mixing calculations of \citet{Howard2023a}, parameterized as
\begin{equation}\label{eq:vmix}
    \Delta V_{\rm mix}(P,T,Y') = \widetilde{V}_{\rm mix}(P,T)\,(1-Y')\,Y'\, ,
\end{equation}
where $ \widetilde{V}_{\rm mix}(P,T)$ is the quantity obtained using the \cite{Howard2023a} table values.

The mixture density is given by $\rho_{\rm mix} = 1/\Delta V_{\rm mix}$. The non-ideal corrections from \cite{Howard2023a} scale as $\sim$$XY$, so we multiply by $Y'(1 - Y')$. Similarly, the specific entropy at a given pressure-temperature coordinate is

\begin{equation}\label{eq:s_val_xy}
    \begin{split}
           S_{XY}(P, T) &= (1 - Y)S_{\rm H}(P, T) + \\[2ex]
           & Y S_{\rm He}(P, T)+ \Delta S_{\rm mix}(P, T) Y (1 - Y)\, .
    \end{split}
\end{equation}
The entropy corrections $\Delta S_{\rm mix}$ already account for the ideal gas entropy of mixing. Likewise, the specific internal energy of H-He mixtures is obtained using 

\begin{equation}\label{eq:u_val}
    \begin{split}
           U_{XY}(P, T) &= (1 - Y)U_{\rm H}(P, T) + \\[2ex]
           & Y U_{\rm He}(P, T)+ \Delta U_{\rm mix}(P, T) Y (1 - Y)\, .
    \end{split}
\end{equation}
Note that the non-ideal corrections of \cite{Howard2023a} are applied only to the CMS19 EOS, not the CD21 EOS, since the CD21 EOS already accounts for these effects in its hydrogen EOS. Recent experiments by \cite{Liu2025} have shown that both the CMS19+HG23 and the CD21 EOSes can reasonably match experimental data, so either EOS should be adequate for evolutionary purposes. 

\begin{figure}[!htb]
\epsscale{1.0}
\plotone{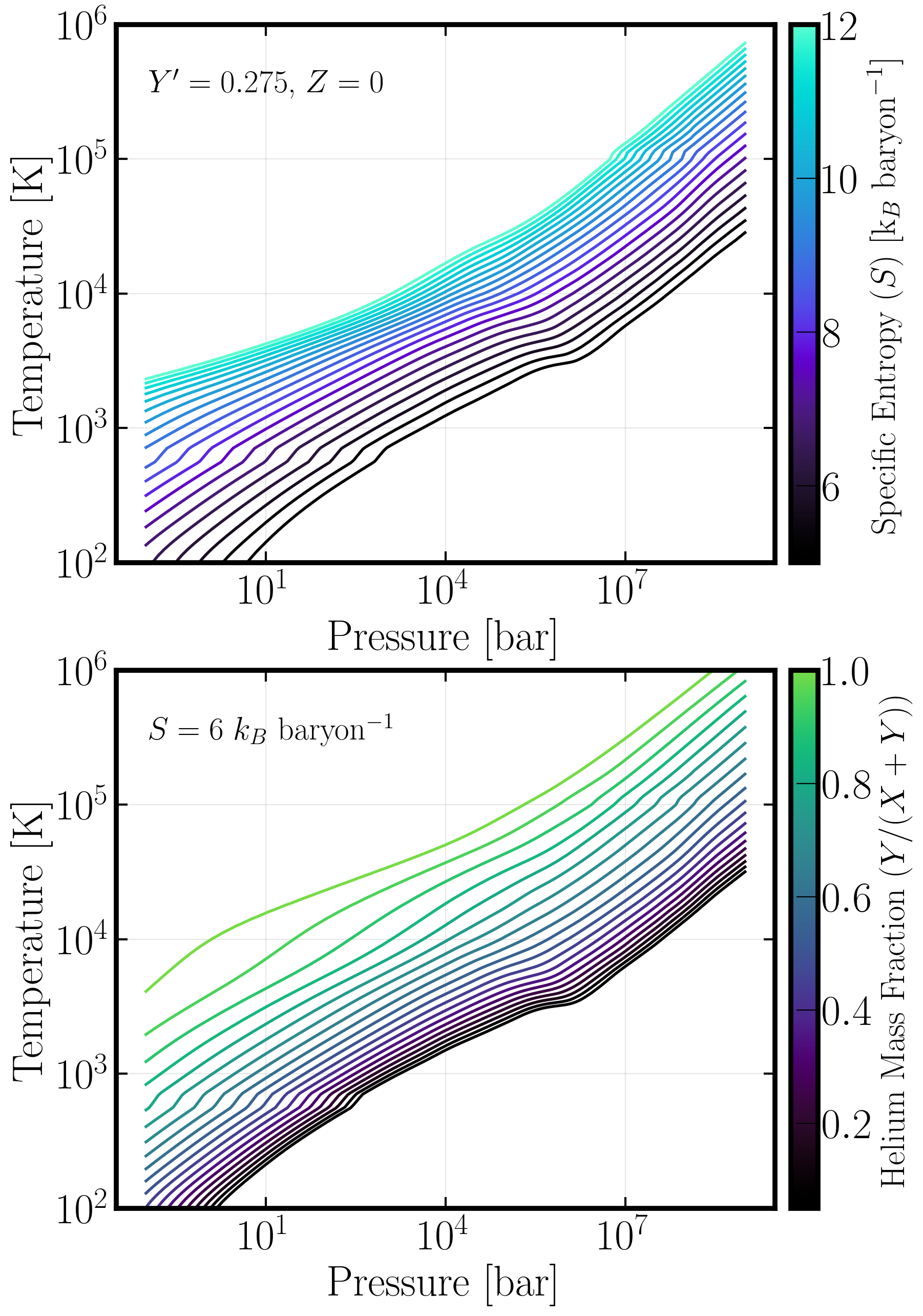}
\caption{Isentropic (constant entropy) temperature profiles of H-He (CD21) mixtures as a function of specific entropy (top), and helium mass fraction (bottom). Isentropic temperatures with higher helium mass fractions are hotter for the same specific entropy. Isentropic temperature profiles are traditionally used to estimate convective regions in envelopes of gas giants or sub-Neptunes. The bumps visible near 800 K are the transitions below $\rho < 0.05$ g cm$^{-2}$ between the \cite{Chabrier2019, Chabrier2021} and \cite{Saumon1995} EOSes. Such transition regions are also visible in Figure 23 of \cite{Chabrier2019}.}
\label{fig:eos_isentropes}
\end{figure}

\begin{figure}[!t]
\plotone{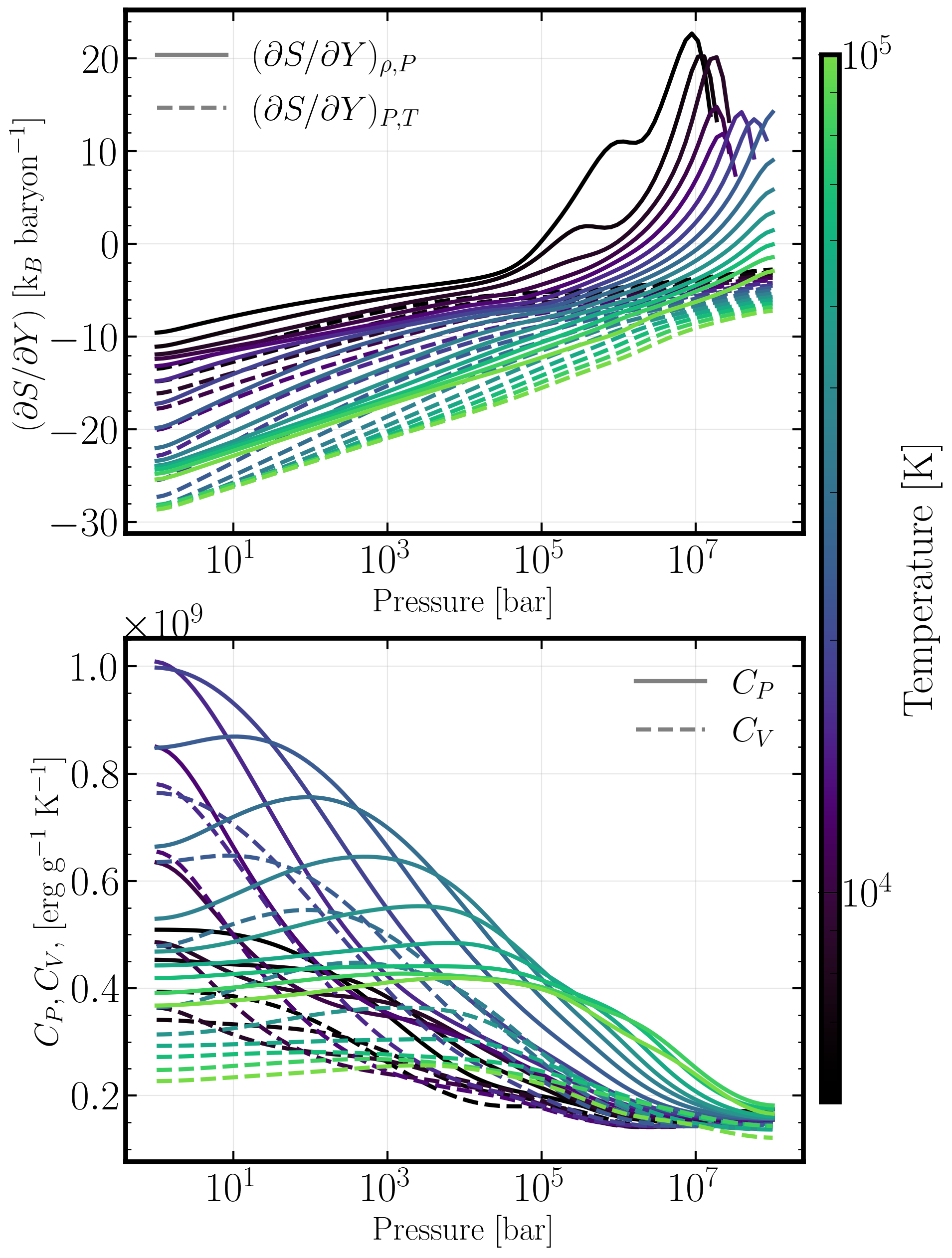}
\caption{Demonstration of CD21 EOS derivatives used in \orchard at $Y/(X+Y) = 0.275$. The top panel shows the entropy-composition derivatives used within the Schwarzschild and Ledoux convective condition (Equation~\ref{eq:ledoux_condition}). The bottom row shows the isobaric and iso-volumetric specific heats in solid and dashed lines, where the former is used in convective flux calculations (Equation~\ref{eq:fconv_apple}). Notably, Equation~\ref{eq:ledoux_condition} can be satisfied when $(\partial S/\partial Y)_{\rho, P} >0$ even when $dY/dr < 0$. Hence, the criterion for convection at $\gtrsim$1 Mbar pressures can be satisfied at $\sim$10$^4$ K even in the presence of a composition gradient. The Schwarzschild condition derivative, $(\partial S/\partial Y)_{P, T}$ (shown in dashed, top), however, remains negative, producing stable conditions when $dY/dr < 0$. The same is true for $Z$ profile gradients.}
\label{fig:eos_derivatives}
\end{figure}

As explained in \cite{Tejada2024}, instead of calculating isentropes by integrating the adiabatic gradient ($\nabla_{\rm ad}$), we compute $S(P, T) \rightarrow T(S, P)$, such that isentropic temperature profiles are enforced by setting a constant $S$ value. We show a range of H-He isentropes in Figure~\ref{fig:eos_isentropes}, and H-He derivatives using the CD21 EOS in Figure~\ref{fig:eos_derivatives}. The top and bottom panel of Figure~\ref{fig:eos_isentropes} shows the temperature profile dependence on the constant entropy and helium mass fraction, respectively. The derivative examples in Figure~\ref{fig:eos_derivatives} are used in calculating the convective flux (Equations~\ref{eq:fconv_apple}, \ref{eq:ledoux_condition}). They are calculated with cell-centered finite differences over a dense grid of precomputed values in a 4-D thermodynamic space such as $S, P, Y, Z$ or $S, \rho, Y, Z$. The entropy-composition derivatives shown in the upper-right panel of Figure~\ref{fig:eos_derivatives}, for example, are computed directly with either the $P, T$ or $\rho, P$ inversions. An inexhaustive list of EOS derivatives and their independent bases is shown in Table 1 of \cite{Tejada2024}. 

\subsubsection{Hydrogen-Helium-Metal Mixture EOSes}\label{subsec:xyz_mix}

We provide H-He-$Z$ mixtures to use in \orchard at any $Z$ fraction. The default metal ($Z$) equation of state used in gas giant and sub-Neptune envelopes is that of water \citet[][``AQUA'']{Haldemann2020}. Users can set this by setting \texttt{z\_eos = aqua} in their configuration parameter file, with options to include the latest AQUA EOS calculated by \cite{Amoros2026}. \orchard also has the option to include the 4:1:7 methane, ammonia, and water solar ratio mixture using the ab initio methane and ammonia EOSes of \cite{Bethkenhagen2017}, coupled with the ammonia EOSes of \cite{Gao2023} and \cite{Setzmann1991} at temperatures lower than 1000 K and densities lower than 0.6 g cm$^{-3}$. This is the EOS used in \cite{Tejada2025} to model the inhomogeneous evolution of Uranus and Neptune. Users can select this by setting \texttt{z\_eos = ice\_mixture} in their configuration file. We also provide H-He mixtures with water and rocks, using a pressure-temperature-dependent Keane EOS \citep{Keane1954} as published and calculated by \cite{Zhang2022}. Users can set this by leaving \texttt{z\_eos = aqua}, setting \texttt{rock\_mixtures = True}, and specifying a rock fraction by setting \texttt{f\_rock\_ini} $\geq 0$. 

To combine these metal equations of state with the H-He EOS at all metal fractions (between 0 and 1.0), we again employ the volume addition law:

\begin{equation}\label{eq:s_val_xyz}
    \begin{split}
           S_{XYZ}(P, T) &= (1 - Z)(S_{XY}(P, T)- S_{\rm mix,id}^{(xy)})\\[2ex] 
           &+ ZS_Z(P, T) + S_{\rm mix,id}^{(xyz)}(1 - Z)\, ,
    \end{split}
\end{equation}
where we subtract the ideal entropy of mixing term due to only helium ($S_{\rm mix,id}^{(xy)}$) and re-add the ideal entropy of mixing term due to helium and metals ($S_{\rm mix,id}^{(xyz)}$).

For a mixture of $N$ molecular species with mass fractions $\{f_i\}$ and molecular weights $\{\mu_i\}$, the ideal entropy of mixing per unit mass (in units of $k_B$ per baryon) is
\begin{equation}\label{eq:smix_id}
    S_{\rm mix,id} = -\frac{1}{\bar{\mu}} \sum_{i=1}^{N} x_i \ln x_i\, ,
\end{equation}
where the number fractions are
\begin{equation}
    x_i = \frac{f_i / \mu_i}{\sum_j f_j / \mu_j}
\end{equation}
and the mean molecular weight is
\begin{equation}
    \bar{\mu} = \sum_{i=1}^{N} \mu_i\, x_i\, .
\end{equation}
Our framework supports up to nine species: H ($\mu=1$), H$_2$ ($\mu = 2$), He ($\mu=4.003$), H$_2$O ($\mu=18.015$), CH$_4$ ($\mu=16.04$), NH$_3$ ($\mu=17.031$), MgSiO$_3$ ($\mu=100.389$), Mg$_2$SiO$_4$ ($\mu = 140.693$) and Fe ($\mu=55.845$). For the special case of a binary H--He mixture, this reduces to
\begin{equation}\label{eq:smix_id_y}
    S_{\rm mix,id}^{(xy)} = -\frac{x_{\rm H}\ln x_{\rm H} + x_{\rm He}\ln x_{\rm He}}{\mu_{\rm H}\,x_{\rm H} + \mu_{\rm He}\,x_{\rm He}}\, ,
\end{equation}
where $x_{\rm He} = Y'/\mu_{\rm He}\,/\,(Y'/\mu_{\rm He} + (1-Y')/\mu_{\rm H})$. This is then generalized to include the ideal entropy-of-mixing terms for the metal species.

Users should note that while H-He non-ideal interactions have been incorporated \citep[see Section~\ref{subsec:xy_mix};][]{Howard2023a}, non-ideal interactions between H-He mixtures and metals have gone relatively unexplored. \cite{Bethkenhagen2017} tested the reliability of the volume addition law compared with real ab initio EOS mixtures of ices and found that the volume addition law was good to within $4\%$. Additionally, \cite{Soubiran2016} investigated ab initio H-He mixtures with C, N, O, Si, Fe, MgO, and SiO$_2$, and found that H-He-$Z$ mixtures computed with the volume addition law remain reliable, but the inclusion of such non-ideal interactions at $\sim$ Mbar pressures remains understudied.  

\subsubsection{Pure Metal EOSes}\label{subsubsec:pure_metal}

\begin{figure}[!htb]
\epsscale{1.0}
\plotone{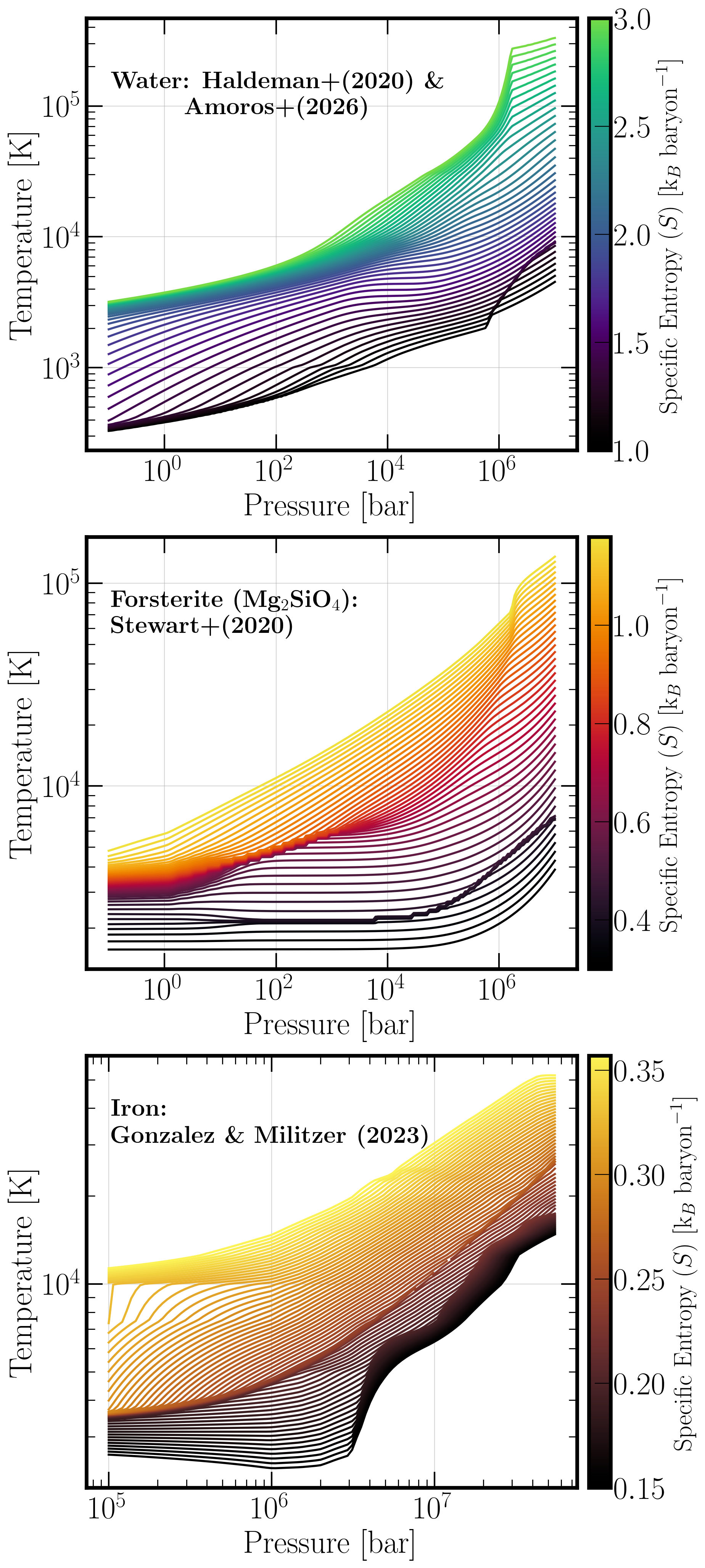}
\caption{Isentropes (constant entropy) along the range of temperatures and pressures of interest of the pure water, AQUA \citep{Haldemann2020} revised EOS of \cite{Amoros2026} (top panel), the \olv\ ANEOS \citep{Thompson1990} EOS of \cite{Stewart2020} (middle panel), and the iron \textit{ab initio} EOS of \cite{Gonzalez-Cataldo2023} (bottom panel). The water EOS can be used in the envelopes with H-He mixtures, while the forsterite and iron EOSes are used in mantles and core regions.}
\label{fig:z_eos_isentropes_demo}
\end{figure}

The design of \orchard assumes that gas giant compact cores are the same as the mantles and cores of smaller planets. As a result, these EOSes are treated the same across all planet types where a $Z=1$ region is specified, such as a compact core in a gas giant or the mantle and core of sub-Neptunes or super-Earths. In \txt{ORCHARD}, the core and mantle masses are user inputs. We provide pure EOS tables for water \citep{Haldemann2020, Mazevet2021, Amoros2026}, methane \citep{Setzmann1991, Bethkenhagen2017}, ammonia \citep{Gao2023, Bethkenhagen2017}, \ppv\, \citep{Luo2025}, \olv\ \citep{Thompson1990, Stewart2020}, iron alloys \citep{Fischer2012}, and pure iron \citep{Ichikawa2014, Dorogokupets2017, Gonzalez-Cataldo2023}. Example isentropes of the revised AQUA table \citep{Amoros2026}, the \olv\, \citep{Stewart2020}, and the iron EOS of \cite{Gonzalez-Cataldo2023} are shown in Figure~\ref{fig:z_eos_isentropes_demo}. 

The default EOS for the mantle of sub-Neptunes, gas giants, and ice giants is Forsterite (Mg$_2$SiO$_4$), with options to use Bridgmanite/perovskite/post-perovskite/enstatite (MgSiO$_3$). The \olv EOS is an updated ``Analytic Equations Of State" (ANEOS) variant, refitted by \cite{Stewart2020} and originally used for shock physics and covering the liquid, solid, and vapor phases. The \ppv EOS combines the ab initio liquid \ppv EOS from \cite{Luo2025}. For the solid phase of \ppv\,  $<$ 23 GPa, we use the solid enstatite Birch-Murnaghan \citep[][(BM3)]{BM3} EOS calculated by \cite{Angel2002}, then for $23 \leq P < 95$ GPa, we use the solid perovskite EOS calculated by \cite{Tange2012}, above which for pressures $\geq 95$ GPa we use the post-perovskite solid EOS of \cite{Sakai2016}. A separate \ppv\ EOS will be included in future work that self-consistently calculates and transitions among the liquid, solid, and vapor phases (J.J. Dong 2026, \textit{in prep}). In addition to \olv\ and \ppv, the user can specify using water as the mantle to model the evolution of ``water-worlds'' \citep{Madhusudhan2021, Luque2022}. Users can set \texttt{mantle\_comp = mg2sio4}, \texttt{mantle\_comp = mgsio3}, or \texttt{mantle\_comp = h2o} in their configuration parameter files to use either of the EOSes described here.

To calculate partially molten mixtures of \olv or \ppv, we use the volume-addition law weighted by the melt fraction (Equation~\ref{eq:melt_fraction}):


\begin{equation}\label{eq:blend_rho}
    \frac{1}{\rho_{\rm blend}} = \frac{1 - \chi}{\rho_{\rm sol}} + \frac{\chi}{\rho_{\rm liq}}\, ,
\end{equation}
where $\rho_{\rm sol}$ is the mass density of the solid phase and $\rho_{\rm liq}$ is the density of the liquid phase. This ensures that the blended specific volume is a linear (and therefore smooth) function of the melt fraction, $\chi$ (Equation~\ref{eq:melt_fraction}). The specific entropy and internal energy are also combined with the volume addition law (see Sections~\ref{subsec:xy_mix} and \ref{subsec:xyz_mix}). The default mantle melt curve of \ppv\ is that of  \cite{Fei2021}:

\begin{equation}\label{eq:tmelt_mgsio3}
    T_{\rm melt}^{(\rm MgSiO_3)}(P) = 6295\left(\frac{P}{140\;\mathrm{GPa}}\right)^{0.317}\;\mathrm{K}\, ,
\end{equation}
while the melt curve of \cite{Presnall1993} is used when the mantle composition is chosen to be \olv:

\begin{equation}\label{eq:tmelt_mg2sio4}
    T_{\rm melt}^{(\rm Mg_2SiO_4)}(P) = 2171\left(1 + \frac{P}{2.44\;\mathrm{GPa}}\right)^{0.088}\;\mathrm{K}\, .
\end{equation}

The EOS for the iron cores of all planets, should the user select an iron core, is that of \cite{Gonzalez-Cataldo2023}. This EOS includes both the liquid and solid hexagonal close-packed (hcp) structures. By default, these settings use the combined EOS via \texttt{core\_comp = Fe\_pure} and \texttt{eos\_core = D17\_comb}.

Using the volume addition law, we calculate the liquid and solid combined EOS across the iron melt curve of \cite{Gonzalez-Cataldo2023}:

\begin{equation}\label{eq:tmelt_fe}
    T_{\rm melt}^{(\rm Fe)}(P) = 6469\left(\frac{P-300}{434.82\;\mathrm{GPa}} + 1\right)^{0.54369}\;\mathrm{K}\,.
\end{equation}
We provide all the solid phases included in \cite{Gonzalez-Cataldo2023} and \cite{Dorogokupets2017} for the user's convenience. These iron EOSes are used for the cores, and users can change them by choosing either \texttt{eos\_core = D17\_comb} or \texttt{eos\_core = G23\_comb} for \cite{Dorogokupets2017} or \cite{Gonzalez-Cataldo2023}, respectively.  The default iron EOS is that of \cite{Gonzalez-Cataldo2023}, set by \texttt{eos\_core = G23\_comb}, and can be changed to that of \cite{Dorogokupets2017} by setting \texttt{eos\_core = D17\_comb}. Moreover, users also have the option to use the liquid EOS from \cite{Ichikawa2014} (\texttt{eos\_core = I14}), and the iron-alloy Fe$_{16}$Si (\texttt{core\_comp = Fe\_alloy}), but these EOSes are available only in the liquid phase. By construction, all iron resides in a pure-Fe core, and all rock/silicate material is in the mantle.\footnote{Generally, chemical phase equilibria between iron and rock materials \citep{Young2024} can also guide ``equilibrium'' abundances corresponding to temperatures and pressures of the mantle and core. These might be included in a later version of \txt{ORCHARD}. Changing the composition also requires EOS mixtures of these materials, and such EOS mixtures need to be pre-tabulated, which requires more computing memory.}

\begin{deluxetable*}{llcl}
\tablecaption{Summary of equations of state available in \orchard.\label{tab:eos_summary}}
\tablecolumns{4}
\tablewidth{\columnwidth}
\tablehead{
\colhead{EOS} &
\colhead{Description} &
\colhead{In H-He mix?\tablenotemark{a}} &
\colhead{References}
}
\startdata
\cutinhead{H--He EOSes (\texttt{hhe\_eos})}
CD21\tablenotemark{$\star$} (\texttt{cd})
    & Envelope
    & \checkmark
    & \citet{Chabrier2021, Militzer2013} \\
CMS19+HG23 (\texttt{cms})
    & Envelope
    & \checkmark
    & \citet{Chabrier2019, Howard2023a} \\
SCvH95 (\texttt{scvh})
    & Envelope\tablenotemark{b}
    & \checkmark
    & \citet{Saumon1995} \\
\cutinhead{Envelope Metal ($Z$) EOSes (\texttt{z\_eos})}
AQUA\tablenotemark{$\star$} (\texttt{aqua})
    & Envelope; H$_2$O
    & \checkmark
    & \citet{Haldemann2020, Mazevet2021} \\
Ice mix\tablenotemark{c} (\texttt{ice\_mixture})
    & Envelope; ices
    & \checkmark
    & \citet{Bethkenhagen2017, Gao2023, Setzmann1991} \\
Rock (Keane)\tablenotemark{d}
    & Envelope; rock
    & \checkmark
    & \citet{Keane1954, Zhang2022} \\
\cutinhead{Mantle EOSes (\texttt{mantle\_comp}; $Z\!=\!1$)}
Mg$_2$SiO$_4$\tablenotemark{$\star$} (\texttt{mg2sio4})
    & Mantle; ANEOS
    & \nodata
    & \citet{Stewart2020} \\
MgSiO$_3$\tablenotemark{e} (\texttt{mgsio3})
    & Mantle; ppv composite
    & \checkmark
    & \citet{Luo2025, Angel2002, Tange2012, Sakai2016} \\
H$_2$O (\texttt{h2o})
    & Mantle; water worlds
    & \checkmark
    & \citet{Haldemann2020, Mazevet2021} \\
MgSiO$_3$ melt
    & Mantle; $\Delta T\!\approx\!50$~K
    & \nodata
    & \citet{Fei2021} \\
\cutinhead{Core EOSes (\texttt{core\_comp}; $Z\!=\!1$)}
Fe liq.+sol.\tablenotemark{$\star$} (\texttt{D17\_comb})
    & Core; liq.+sol.\ (hcp)
    & \nodata
    & \citet{Dorogokupets2017} \\
Fe liq.\ only (\texttt{I14})
    & Core; liquid only
    & \nodata
    & \citet{Ichikawa2014} \\
Fe$_{16}$Si (\texttt{Fe\_alloy})
    & Core; liquid only
    & \nodata
    & \citet{Fischer2012} \\
Fe melt
    & Core; $\Delta T\!=\!50$~K
    & \nodata
    & \citet{Zhang2015} \\
\enddata
\tablenotetext{\star}{Default option in \orchard.}
\tablenotetext{a}{Indicates whether the EOS enters the volume-addition H-He-$Z$ mixing framework (Section~\ref{subsec:xyz_mix}). Mantle and core EOSes operate in $Z=1$ regions.}
\tablenotetext{b}{Also used as the low-density ($\rho < 0.05$~g~cm$^{-3}$) supplement to CMS19.}
\tablenotetext{c}{4:1:7 CH$_4$:NH$_3$:H$_2$O solar-ratio mixture.}
\tablenotetext{d}{Enabled via \texttt{z\_eos = aqua}, \texttt{rock\_mixtures = True}.}
\tablenotetext{e}{Solid phases: enstatite BM3 ($<$23~GPa), perovskite (23--95~GPa), post-perovskite ($\geq$95~GPa). Liquid: ab initio.}
\tablecomments{EOS tables described in Section~\ref{subsec:EOS}, separated by where each EOS is applicable in the interiors of all planets \orchard is designed to model.}
\end{deluxetable*}

\subsection{Atmospheres}\label{subsec:atm}

The first atmosphere models for evolutionary calculations were published by \cite{Graboske1975}. \cite{Hubbard1977, Hubbard1978}, and \cite{Pollack1977} applied them to the first thermal cooling models of Jupiter and Saturn coupled to boundary conditions. As discussed in \papI, \cite{Burrows1997} computed the first non-gray 1-D atmospheric models for gas-giant and brown-dwarf thermal evolution. \orchard inherits this method, wherein tables from a radiative transfer code, such as \texttt{CoolTLusty} \citep{HubenyLanz1995, Sudarsky2003, Sudarsky2005, Burrows2008, Hubeny2011}, are interpolated to derive internal flux temperatures ($T_{\rm int}$) and surface losses. We incorporate such flux boundary conditions for all planet masses of interest and describe these in the following sections, summarized in Table~\ref{tab:atm_summary}. 

\subsubsection{Solar System Atmosphere Models}

For our own published Jupiter and Saturn boundary conditions with irradiation and ammonia clouds \citep[][C23]{Chen2023}, we use grids of interior entropy value ($S_{\rm int}$) as a function of $\Tint$, surface $\log_{10}{g}$, and helium mass fraction, $Y$. We then inverted these to derive $\Tint$ as a function of the interior $S_{\rm int}$ and $\log_{10}{g}$, and interpolated the result during an evolutionary run. The C23 boundary conditions connect to the evolution model at an average pressure of 10--30 bars. The incident instellation is fixed to match that of the Sun at the corresponding distances. The methodology for solar irradiation is described in \cite{Chen2023}.  To select these boundary conditions, the user must set \texttt{bc\_atm = c23} and set \texttt{planet = Jupiter} or \texttt{planet = Saturn}. These tables span a range of $\log_{10}{g} \in [2.4, 3.6]$ and $\Tint \in [90, 450]$ K. For all boundary conditions, the intrinsic flux is calculated with the Stefan-Boltzmann law,

\begin{equation}\label{eq:sb_law}
    \mathcal{F_{\rm int}} = \sigma_{\rm SB}\Tint^4\, ,
\end{equation}
where $\sigma_{\rm SB} = 5.6704\times10^{-8}$ W m$^{-2}$ K$^{-4}$ is the Stefan-Boltzmann constant.

Furthermore, we make available the Jupiter, Saturn, Uranus, and Neptune boundary conditions of \cite{Fortney2011}. To use each of these boundary conditions, the user must set \texttt{bc\_atm = f11} and select the planet accordingly, e.g., \texttt{planet = Saturn} or \texttt{planet = Neptune}. These boundary conditions use the temperature at 10 bars, $\Tten$, mapped onto a 3-D grid of $\log_{10}{g}$ and $\Tint$, and assume insolation at either 1.0 or 0.7 solar luminosity. The $\log_{10}{g}$ values provided are $\in [2.5, 3.45]$ for Jupiter, $\in [2.1, 3.1]$ for Saturn, and $\in [2.5, 3.1]$ for Uranus and Neptune. These boundary conditions connect to the evolution model using the temperature at 10 bars. 

In addition to these newer boundary conditions, we also include in \orchard the older fits to Uranus and Neptune boundary conditions calculated by \cite{Graboske1975} and \cite{Hubbard1977}, relating $\Teff$, the temperature at 1 bar ($T_{\rm 1 bar}$), and the surface gravitational acceleration, $g$:

\begin{equation}\label{eq:Tint_g75}
    T_{\rm eff} = \bigg(\frac{T_{\rm 1 bar}}{K} g^{1/6}\bigg)^{1/1.244}\, ,
\end{equation}
where $T_{\rm 1 bar}$ is the temperature at 1 bar provided by the interior model, and $g = GM/R^2$. Here, $K$ is a fitted value corresponding to $K = 1.481$ for Uranus, and $K = 1.451$ for Neptune \citep{Graboske1975, Scheibe2019}. The reader should keep in mind that Equation~\ref{eq:Tint_g75} is a fitted relation, not a physically derived relation. Then, the internal temperature is obtained via

\begin{equation}\label{eq:Teff}
    T_{\rm int} = (T_{\rm eff}^4 - T_{\rm eq}^4)^{1/4}\, ,
\end{equation}
where $T_{\rm eq}$ is the equilibrium temperature:

\begin{equation}\label{eq:Teq}
    T_{\rm eq} = T_\odot(1 - A)^{1/4} \bigg(\frac{R_\odot}{2d}\bigg)^{1/2}\, ,
\end{equation}
where $A$ is the Bond albedo, $d$ is the orbital semi-major axis, and $T_\odot$ and $R_\odot$ are the effective temperature and radius of the Sun, respectively. The Bond albedos of 0.3 and 0.29 \citep{Pearl1991}, and $d = 19.19$ A.U. and $d = 30.07$ A.U. in Equation \ref{eq:Teq}, respectively, as discussed in \cite{Scheibe2019} and \cite{Tejada2025b}. Since these are analytic fits, they could, in principle, apply to any range of temperatures and gravitational accelerations. To select this boundary condition, the user must set \texttt{bc\_atm = g75} and \texttt{planet = Uranus} or \texttt{planet = Neptune}. The input parameters to the Uranus and Neptune atmosphere models are the Bond albedo, $K$, and the equilibrium temperature. The calculated values from the interior structure model are the gravitational acceleration $g$, and the temperature at 1 bar. 

\subsubsection{Gas Giant Exoplanet Atmosphere Models}

To model a wider range of gas giant exoplanet masses, we incorporate updated atmospheric boundary conditions for isolated, cloud-free objects \citep[][C26]{Chen2026}. These atmosphere models were calculated with \texttt{CoolTLusty} \citep{HubenyLanz1995, Sudarsky2000, Sudarsky2003, Sudarsky2005, Burrows2008, Hubeny2012}, employing updated atmosphere opacity calculations from \cite{Lacy2023}. These boundary conditions span $\log_{10}{g} \in [2.8, 4.4]$ and $\Tint \in [100, 1400]$ K. These boundary conditions are calculated at two helium fractions $Y= 0.15, 0.275$ and three metal fractions, $Z = 0.017,0.051, 0.145$ for 1, 3.16, and 10 times the solar abundance (assuming the solar abundance, $Z_{\odot}$, is 0.017). We then interpolate across these sheets to obtain the general $\Tint, \Teff$ dependence on $\log_{10}{g}$, $S_{\rm int}$, $Y$, and $Z$. To select this boundary condition, users must specify only \texttt{bc\_atm = c26}. Just as in the C23 atmospheres, we use the entropy of the interior profile at a mean pressure of between 10 and 30 bars to connect to the interior evolution model.

Additional boundary conditions tailored for giant-planet evolution and not including instellation include the Sonora-Bobcat suite of models from \cite{Marley2021}\footnote{These atmosphere models also use $\Tten$ as input. The ranges are: $\log_{10}{g}\in[3,5.5]$, $\Teff \in [200, 2400]$ K, and offer metallicities from 0.316, 1.0, and 3.16 times solar (M/H$ = -0.5$, M/H$ = 0.0$, and M/H$ = 0.5$). We interpolate across these to offer metallicity dependence, $\Teff(\log_{10}{g}, \Tten, Z)$. To select this boundary condition, users must only specify \texttt{bc\_atm = m21}.}, the atmosphere boundary conditions of \citet[][ATMO2020]{Phillips2020}\footnote{where only $\Teff(\log_{10}{g}, \Tten)$ is provided by using \texttt{bc\_atm = p20}. The ATMO2020 parameter ranges are: $\log_{10}{g}\in[2.5,5.5]$, $\Teff \in [200, 3000]$ K.}, and the boundary conditions of \cite{Burrows1997}\footnote{where only $\Teff(\log_{10}{g}, \Tten)$ are provided and one should use \texttt{bc\_atm = b97}. The \cite{Burrows1997} parameter ranges are $\log_{10}{g}\in[2.0,5.5]$ and $\Tten \in [50, 5050]$ K.}.

\subsubsection{Sub-Neptune and Super-Earth Exoplanet Atmosphere Models}

Sub-Neptune atmosphere models differ from those of gas-giant exoplanets in that they may have higher metallicities and be highly irradiated \citep[e.g., GJ1214; b][]{Kempton2023a}. By default, we adopt the non-gray radiative-convective equilibrium atmosphere model of \cite{Ohno2023}, with heritage work from \cite{Fortney2007, Fortney2020}, and updated by \cite{Chachan2025} and \cite{Tang2025}. These atmosphere models yield the internal temperature dependence $\Tint(\log_{10}{g}, T_{1000}, \Teq, Z_{1000}, \mathcal{F}_*)$, where $T_{1000}$ is the temperature at 1 kbar, $Z_{1000}$ is the metal mass fraction at 1000 bars, and $\mathcal{F}_*$ is the instellation flux. The planetary interior structure provides $\log_{10}{g},\ T_{1000}$, and $Z_{1000}$. The ranges are: $\log_{10}{g} \in [1.0, 5.0]$, $T_{1000} \in [100, 10^4]$ K, $Z_{1000} \in [1, 100]\ Z_\odot$, and $\mathcal{F}_* \in [0.73, 1000]\, F_\oplus$, where $F_\oplus = 1361$ W m$^{-1}$. After calculating $\Tint(\log_{10}{g}, T_{1000}, \Teq, Z_{1000}, \mathcal{F}_*)$, the effective temperature is calculated by re-arranging Equation~\ref{eq:Teff}. The incoming instellation flux is calculated assuming a black-body where $\mathcal{F}_* = 4 \pi \sigma \frac{\Teq^4}{1 - A}$, where $A$ is the Bond albedo. Users can specify the $\Teq$ and Bond albedos in their configuration parameter files. For example, \texttt{T\_eq = 400}, \texttt{bond\_albedo = 0.5} after setting \texttt{bc\_atm = f07}.

For rocky planets retaining a thin atmosphere, we adopt an
Eddington approximation temperature--optical depth relation evaluated at the
photosphere \citep[see also][for a detailed discussion.]{Guillot2010}, $\tau_{\rm match} = 2/3$:
\begin{equation} \label{eq:gray_tsurf}
  T_{\rm surf}^{4} = T_{\rm eq}^{4}
    + \frac{3}{4}\,T_{\rm int}^{4}
      \left(\tau_{\rm match} + \frac{2}{3}\right)
    = T_{\rm eq}^{4} + T_{\rm int}^{4},
\end{equation}
so that inverting for the intrinsic temperature yields
\begin{equation} \label{eq:gray_tint}
  T_{\rm int} = \left(
    T_{\rm surf}^{4} - T_{\rm eq}^{4}
  \right)^{1/4}\!,
\end{equation}
and the effective temperature is solved for by re-arranging Equation~\ref{eq:Teff}. We remind the reader that Equation~\ref{eq:Tint_g75} is only for Uranus and Neptune, not the gray atmosphere model described in this section. The default gray infrared opacity is $\kappa_{\rm IR} = 10^{-3}$~cm$^{2}$\,g$^{-1}$.
Users can select this option for either super-Earths or sub-Neptune models by setting \texttt{bc\_atm = gray} and selecting the $\kappa_{\rm IR}$ value: \texttt{rock\_gray\_kappa\_ir = 1e-3}. The purpose of $\kappa_{\rm IR}$ can therefore be used to throttle the escaping luminosity. The input parameters for the \txt{gray} atmosphere model are $\Teq$ and the albedo, $A$. The known parameters from the interior structure model are the gravitational acceleration, $g$, and $T_{\rm surf}$. From here, $\Tint$ is calculated via Equations~\ref{eq:gray_tint}, and $\Teff$ is calculated from Equation~\ref{eq:Teff}.

For planets with no atmospheres at all (bare), the structure is allowed to cool directly into space and hence the effective temperature equals the surface temperature, $T_{\rm eff} = T_{\rm surf}$. $\Tint$ is then obtained from Equation~\ref{eq:Teff}, and the intrinsic flux is obtained with Equation~\ref{eq:sb_law}.Users can use this for super-Earth models (where the total mass is the core mass) and set \texttt{bc\_atm = bare}. This method has been applied for super-Earth evolution models in \seI. A comprehensive summary of the atmosphere models used in \orchard along with their physical ranges is found in Table~\ref{tab:atm_summary}.

\begin{deluxetable*}{lllc}
\tablecaption{Summary of Atmospheric Boundary Conditions in \orchard.\label{tab:atm_summary}}
\tablecolumns{4}
\tablewidth{\textwidth}
\tablehead{
\colhead{Model Name} & 
\colhead{Planet Type} & 
\colhead{Config Parameters} & 
\colhead{Physical Ranges}
}
\startdata
\cutinhead{Solar System Models}
C23\tablenotemark{a,$\star$} (\texttt{c23}) & 
Jupiter/Saturn & 
\texttt{bc\_atm = c23}, \texttt{planet = Jupiter/Saturn} & 
$\log_{10}g \in [2.4, 3.6]$, $T_{\rm int} \in [90, 450]$~K \\
F11\tablenotemark{b} (\texttt{f11}) & 
All Solar System & 
\texttt{bc\_atm = f11}, \texttt{planet = Jup/Sat/Ura/Nep} & 
$\log_{10}g \in [2.1, 3.45]$, $T_{\rm int} \in [50, 300]$~K \\
G75\tablenotemark{c} (\texttt{g75}) & 
Uranus/Neptune & 
\texttt{bc\_atm = g75}, \texttt{planet = Uranus/Neptune} & 
Analytic \\
\cutinhead{Exoplanet Models}
C26\tablenotemark{d, $\ddag$} (\texttt{c26}) & 
Gas Giants & 
\texttt{bc\_atm = c26} & 
$\log_{10}g \in [2.8, 4.4]$, $T_{\rm int} \in [100, 1400]$~K \\
M21\tablenotemark{e} (\texttt{m21}) & 
Gas Giants & 
\texttt{bc\_atm = m21} & 
$\log_{10}g \in [3, 5.5]$, $T_{\rm eff} \in [200, 2400]$~K \\
P20\tablenotemark{f} (\texttt{p20}) & 
Gas Giants & 
\texttt{bc\_atm = p20} & 
$\log_{10}g \in [2.5, 5.5]$, $T_{\rm eff} \in [200, 3000]$~K \\
B97\tablenotemark{g} (\texttt{b97}) & 
Gas Giants & 
\texttt{bc\_atm = b97} & 
$\log_{10}g \in [2.0, 5.5]$, $T_{10} \in [50, 5050]$~K \\
F07\tablenotemark{h,$\dagger$} (\texttt{f07}) & 
Sub-Neptunes & 
\texttt{bc\_atm = f07}, \texttt{T\_eq}, \texttt{bond\_albedo} & 
$\log_{10}g \in [1.0, 5.0]$, $T_{1000} \in [100, 10^4]$~K \\
Gray\tablenotemark{i} (\texttt{gray}) & 
Rocky/Sub-Nep & 
\texttt{bc\_atm = gray}, \texttt{rock\_gray\_kappa\_ir} & 
Analytic \\
Bare\tablenotemark{j} (\texttt{bare}) & 
Airless & 
\texttt{bc\_atm = bare} & 
Analytic \\
\enddata
\tablenotetext{\star}{Default for Solar System Jupiter/Saturn evolution.}
\tablenotetext{\ddag}{Default for gas giant exoplanets.}
\tablenotetext{\dagger}{Default for sub-Neptunes.}
\tablenotetext{a}{\citet{Chen2023}}
\tablenotetext{b}{\citet{Fortney2011}}
\tablenotetext{c}{\citet{Graboske1975, Hubbard1977, Scheibe2019}}
\tablenotetext{d}{\citet{Chen2026}}
\tablenotetext{e}{\citet{Marley2021}}
\tablenotetext{f}{\citet{Phillips2020}}
\tablenotetext{g}{\citet{Burrows1997}}
\tablenotetext{h}{\citet{Ohno2023, Fortney2007, Chachan2025, Tang2025}}
\tablenotetext{i}{Equations \ref{eq:gray_tsurf}--\ref{eq:gray_tint}}
\tablenotetext{j}{\citet{Zhang2022}}
\tablecomments{Summary of atmospheric boundary conditions across the planetary scale of interest. Parameters in \texttt{monospaced} text refer to configuration flags. Physical ranges list the bounds of the lookup tables; analytic models are applicable across all relevant parameter spaces. The atmosphere boundary condition module in \orchard is called \href{https://github.com/robtejada/orchard/blob/main/atm_bc.py}{\txt{atm\_bc.py}}.}
\end{deluxetable*}

\section{Example Calculations \& Applications}\label{sec:examples}

\begin{figure*}[!t]
\epsscale{1.0}
\plotone{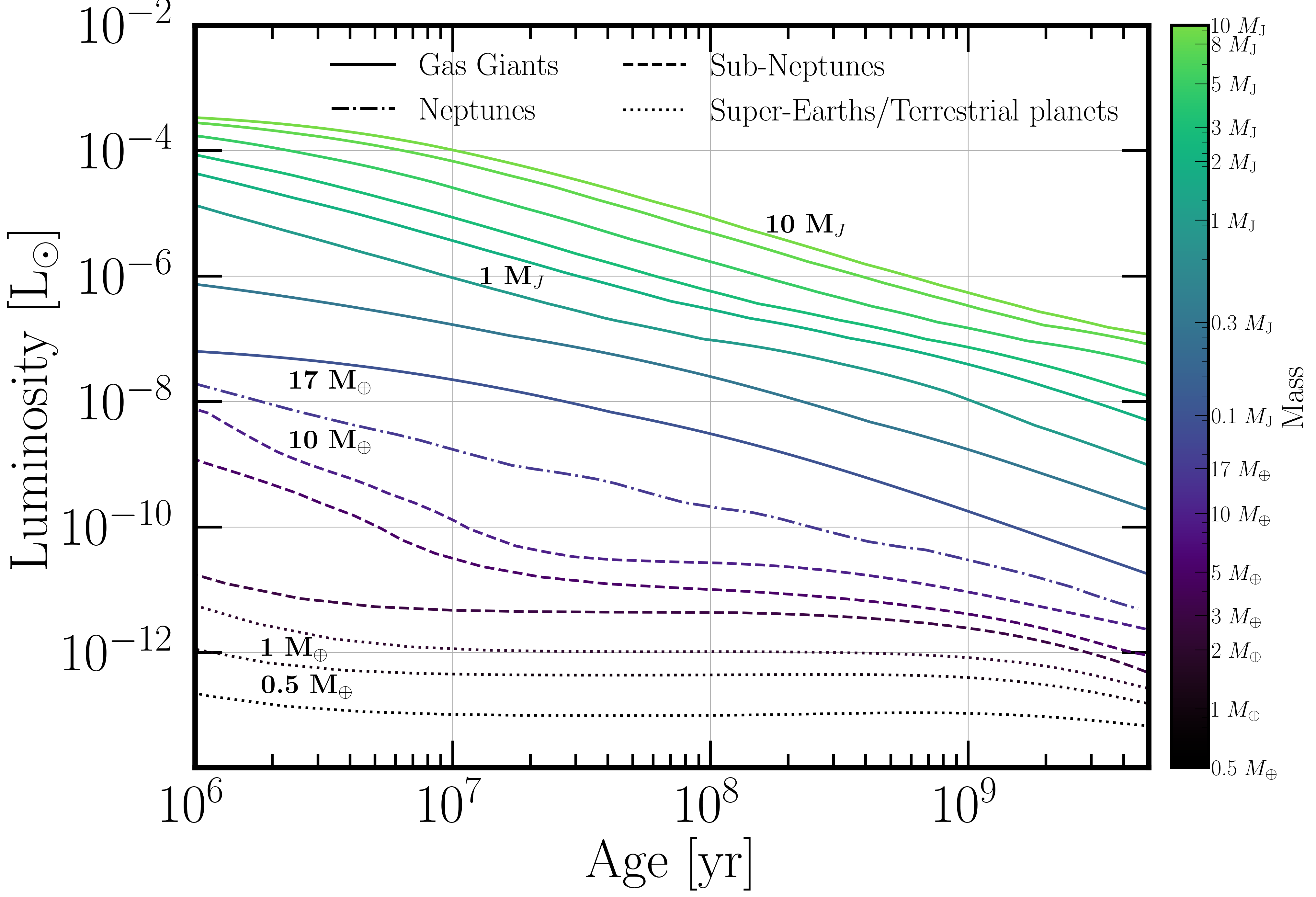}
\caption{Luminosity evolution across the \orchard planetary modeling capability. In this figure, the gas giants follow the total metal content predicted by the mass-metallicity relation of \cite{Chachan2025}, and their atmospheres are calculated using our updated atmosphere models \cite{Chen2026}. The Neptunian planet (17 \mearth) is a homogeneous model at 175 times the solar metallicity, and the three sub-Neptunes (3, 5, and 10 \mearth) contain 40 times the solar metallicity, with 10\% envelope mass fractions. The three super-Earth/terrestrial planets (0.5, 1, and 2 \mearth) have 1\% H-He envelopes. This figure is inspired by Figure 7 of \cite{Burrows1997}, which shows the luminosity evolution of stars, brown dwarfs, and planets down to a Saturn-mass object.}
\label{fig:burrows_97}
\end{figure*}

\begin{figure}[!t]
\epsscale{1.05}
\plotone{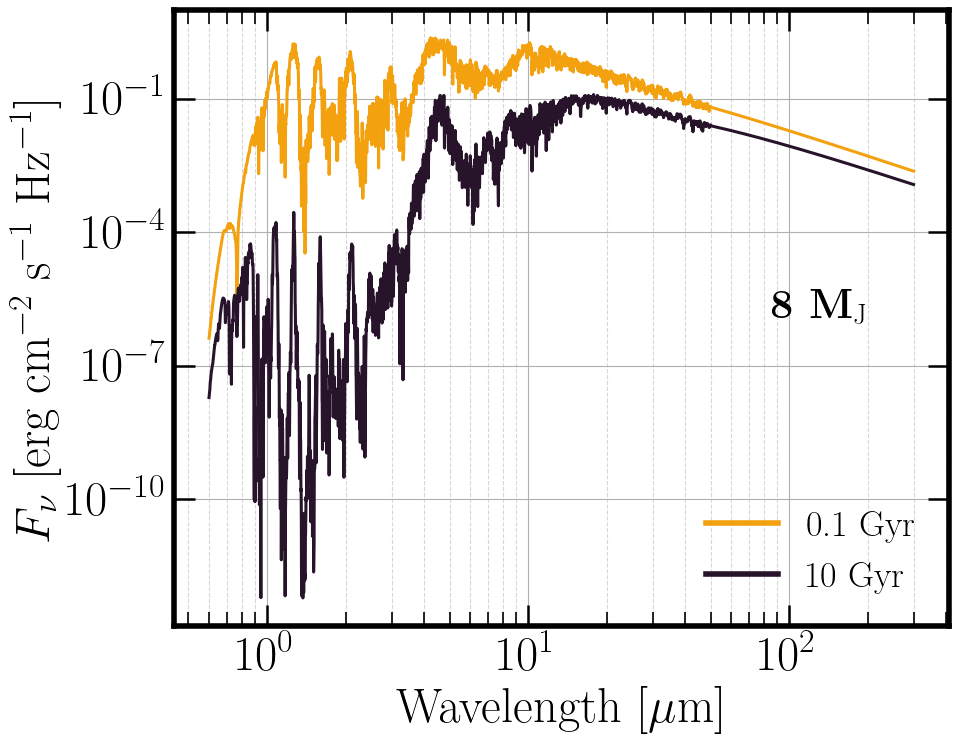}
\caption{Atmospheric spectra evolution of the flux density as a function of wavelength of the 8 \mjup\, model shown in Figure~\ref{fig:burrows_97} with \texttt{ORCHARD}. The spectra are calculated as post-processes using the methods described in \citet[][]{Chen2026}. As such, \orchard is equipped to model atmospheric spectra for direct-imaging investigations of exoplanets.}
\label{fig:spectrum_evol}
\end{figure}

\orchard is designed to meet the demands of Solar System giant and ice giant planet evolution, as well as general exoplanet interior evolution. We show here some example applications of Solar System homogeneous and inhomogeneous models, gas-giant exoplanets, sub-Neptunes, super-Earths, and terrestrial evolution models. First, we show a family of models in Figure~\ref{fig:burrows_97} spanning 0.5 \mearth---10 \mjup. Figure~\ref{fig:burrows_97} was inspired by Figure 7 of \cite{Burrows1997}, where they showed the luminosity evolution of low-mass stars, sub-stellar objects, and gas giant planets down to a Saturn mass. We expand this here to cover Neptunian (dash-dot), sub-Neptunian (dashed), and terrestrial luminosity evolution (dotted) to present a general overview of the scope of \texttt{ORCHARD}. Given how central evolution models are in interpreting direct-imaging observations of gas giant exoplanets, we provide in Figure~\ref{fig:spectrum_evol} a demonstration of atmospheric spectrum evolution for the 8~\mjup\, model shown in Figure~\ref{fig:burrows_97} using the calculations of \citet[][; C26]{Chen2026}. Users can access these spectra tables with \texttt{atm\_spectra/spectrum.py}.

\begin{figure}[!t]
\epsscale{1.1}
\plotone{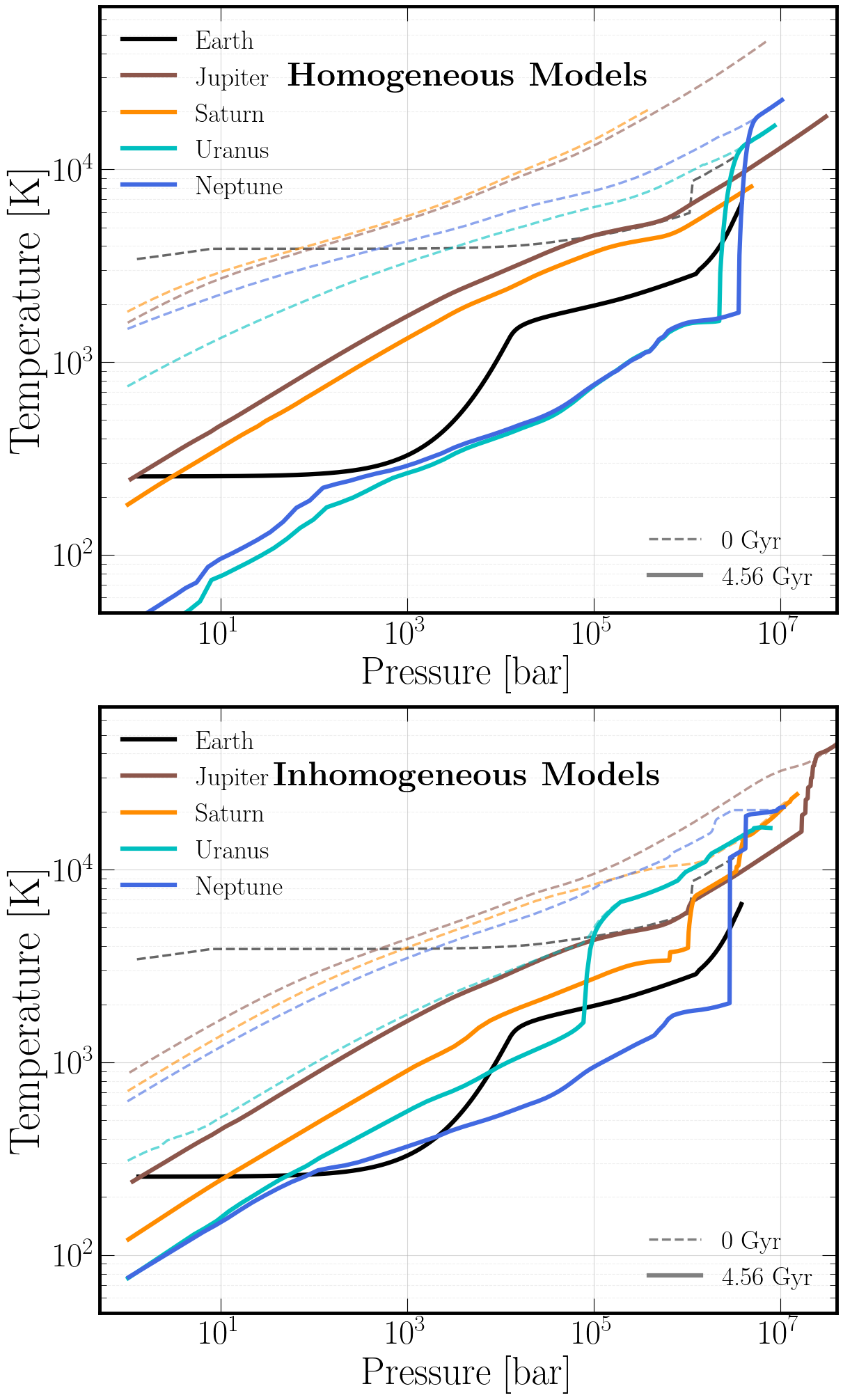}
\caption{Demonstration of Solar-System gas-giant planet evolution, along with a bare (no atmosphere) 1 \mearth mass model (in black). The initial conditions for all models are shown as thin dashed lines in their corresponding colors. The top panel shows examples of traditional homogeneous envelope evolution models for Jupiter, Saturn, Uranus, and Neptune at 3, 5, 100, and 175 times solar metallicity, respectively. The bottom panel shows the inhomogeneous evolution models of the gas giant planets presented in \cite{Tejada2025, Sur2025a} and \cite{Tejada2025b}, comparing two different epochs. The Uranus and Neptune homogeneous-envelope models host 4 \mearth sized rocky cores. The inhomogeneous envelope model paradigm has superseded the traditional homogeneous envelope model of the Solar System gas giants.}
\label{fig:solar_system}
\end{figure}

In Figure~\ref{fig:solar_system}, we present a comparison of homogeneous envelope (top) and inhomogeneous envelope (bottom) models of the Solar System gas and ice giants, along with a fiducial 1~\mearth mass model shown in black. The homogeneous models of Jupiter and Saturn are coreless, making them fully homogeneous in composition. The initial thermal states are shown in thin dashed lines, and the present-age structures are shown in solid lines. The inhomogeneous Solar System models in the bottom panel are versions of those presented in \cite{Tejada2025}, \cite{Sur2025a}, and \cite{Tejada2025b}.\footnote{We provide a description of homogeneous, inhomogeneous, and rocky/iron interior initial conditions in Appendix~\ref{subsec:initial_setup}.} Compared with homogeneous models, inhomogeneous models generally remain hotter in their deep interiors ($\gtrsim 1$ Mbar). This feature is accentuated in the Uranus model, whose intrinsic measured flux can be explained by a large compositional gradient that prevents efficient adiabatic cooling of its interior \citep{Nettelmann2013a, VazanHelled2020, Tejada2025}.

\begin{figure*}[!t]
\epsscale{1.15}
\plotone{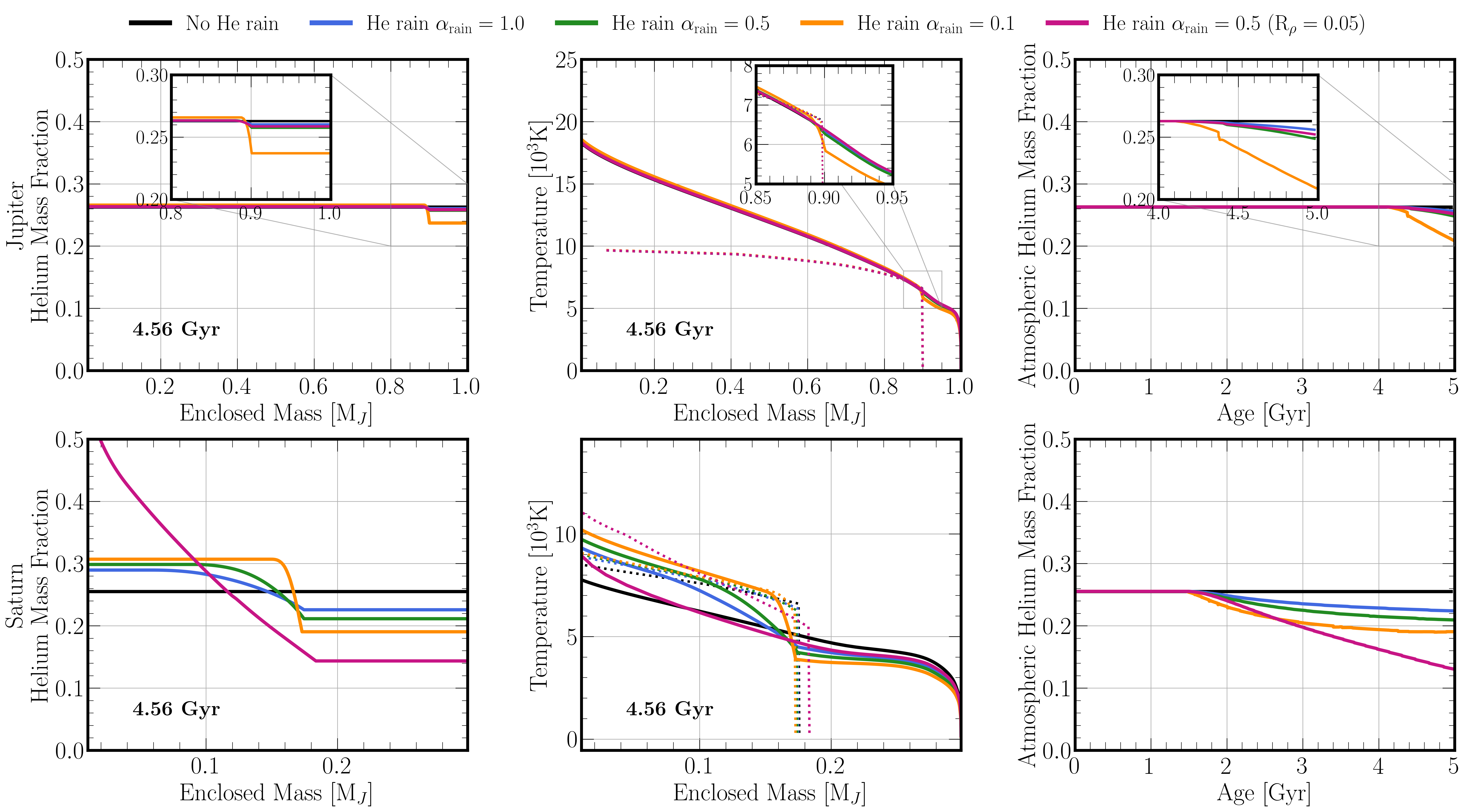}
\caption{Demonstration of initially homogeneous Jupiter and Saturn models undergoing various degrees of helium rain with respect to the helium rain mixing length parameter, $\alpha_{\rm rain}$, compared at 4.56 Gyr. The top row shows the helium, temperature, and atmospheric helium fraction vs. time for the Jupiter model, and the bottom row does the same for Saturn. The Jupiter and Saturn models have heavy-element abundances 3 and 5 times solar throughout their interiors, respectively. The miscibility curve used is that of \cite{Lorenzen2009, Lorenzen2011} (shown as dotted lines).The vertical temperature offset often used by modelers \citep[e.g.,][]{Pustow2016, Nettelmann2015, Mankovich2016} represents higher H-He miscibility temperatures relative to the original curves. We find a shift to higher temperatures by +300 K, as reported by \cite{Sur2025a}. More evident in the Saturn panels (bottom), smaller $\alpha_{\rm rain}$ parameters decrease the pressure scale height used to calculate the sedimentation velocity of the helium in the helium rain regions, so smaller values yield more vigorous rain and, thus, steeper helium gradients (left panel), steeper super-adiabatic temperature profiles (middle panel), and more helium depletion (right panel). The same is shown in the inset panels for Jupiter. The black, blue, green, and yellow models all assume full Ledoux convection, or an $R_\rho = 1.0$. The green model ($\alpha_{\rm rain}$ = 0.5) is contrasted here with a model having $R_\rho = 0.05$ \citep[see][for a detailed discussion]{Sur2025b}. The purple Saturn model experiences more helium depletion and produces a steeper helium rain gradient, creating a so-called ``helium ocean'' \citep[e.g., as reported by][]{Howard2024}. This model is also calculated with a +300 K shift, using the miscibility curves from \cite{Lorenzen2009, Lorenzen2011}. However, assuming that $R_\rho = 0.05$ implies a Schwarzschild-like convective criterion, and is, thus, likely not applicable in the presence of compositional gradients. Such a low $R_\rho$ value, however, yields the classical result of explaining Saturn's luminosity with helium rain heating, as demonstrated in Figure~\ref{fig:sat_lum_demo}. \orchard can thus replicate previous helium rain results. }
\label{fig:jup_sat_he_rain}
\end{figure*}

\begin{figure}[!t]
\plotone{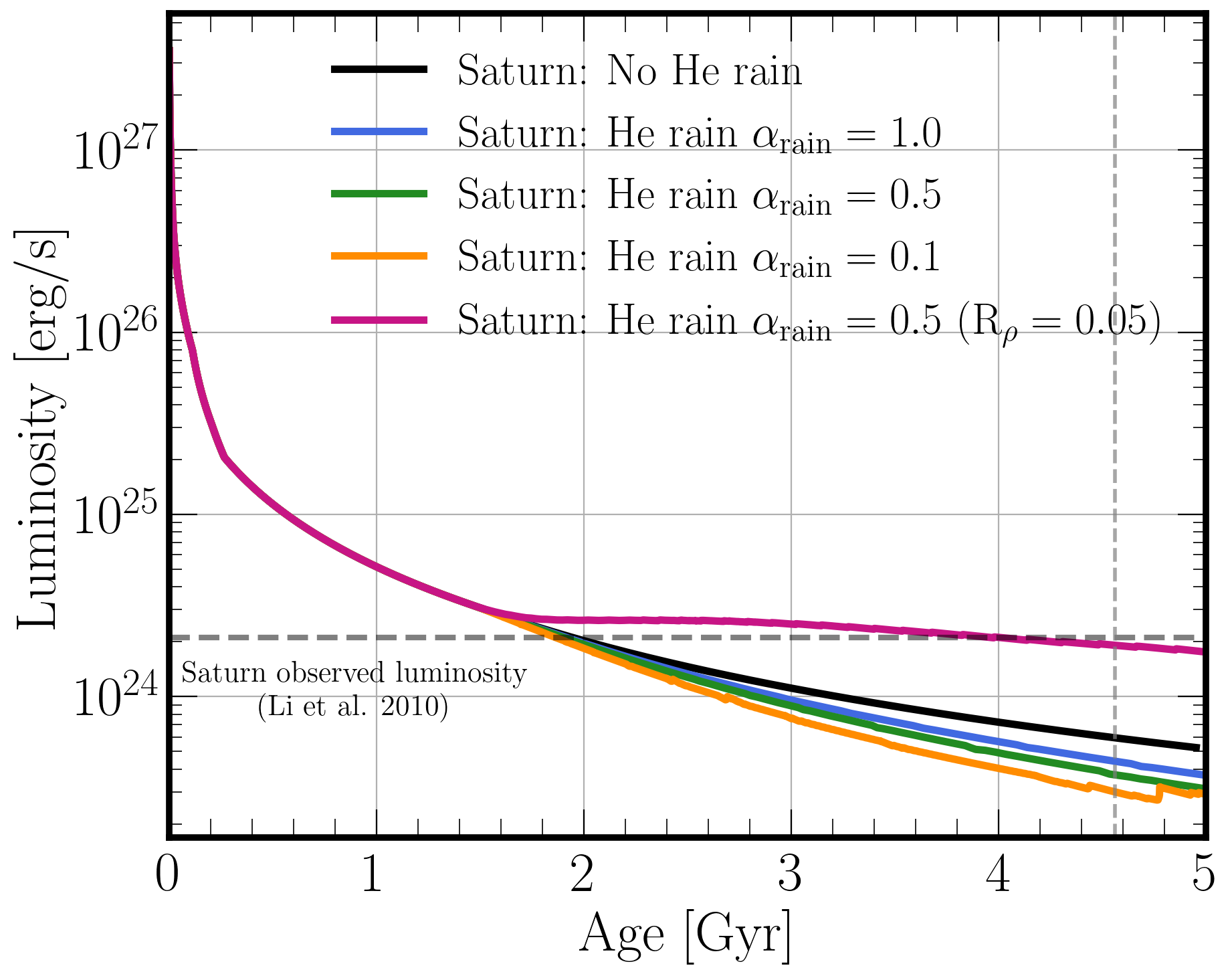}
\caption{Luminosity evolution of the initially homogeneous Saturn models shown in the bottom row of Figure~\ref{fig:jup_sat_he_rain}. The blue, green, and yellow models have smaller helium rain scale height parameters ($\alpha_{\rm rain}$; Equation~\ref{eq:v_sed}) and use the Ledoux criterion for convection ($R_\rho = 1.0$). The black model does not include helium rain. On the other hand, the purple model uses $R_\rho = 0.05$, closer to Schwarzschild convection \citep[e.g.,][]{Mankovich2020} and reproduces the classical result of helium rain heating to explain Saturn's present-day luminosity \citep{Li2010}, shown by the faint dashed lines. See \cite{Sur2025b} for a detailed discussion on the $R_\rho$ parameter.}
\label{fig:sat_lum_demo}
\end{figure}

We demonstrate the evolution of two initially homogeneous models of Jupiter (top row) and Saturn (bottom row) in Figures~\ref{fig:jup_sat_he_rain}. These models are representative of how evolution models were conducted before the \textit{Juno} and \textit{Cassini} fuzzy core results and were aimed at explaining the atmospheric helium abundances, effective temperatures, radii, and luminosities of Jupiter and Saturn. Figure~\ref{fig:jup_sat_he_rain} shows that a smaller helium rain mixing scale parameter, $\alpha_{\rm rain}$ (Equations~\ref{eq:misc} and \ref{eq:v_sed}), yields more vigorous helium rain, resulting in thinner helium rain regions (left column panels) and more helium depletion (right column panels). The temperature profiles align with the miscibility curve (dotted lines, center-column panels), creating steep, super-adiabatic gradients with smaller $\alpha_{\rm rain}$ values. All models except for the purple model use the Ledoux condition (Equation~\ref{eq:ledoux_condition}) for convection, which is equivalent to $R_\rho = 1.0$ in Equation 26 of \papI\, \citep[see][for a detailed discussion]{Sur2025b}. The purple model, on the other hand, uses $R_\rho = 0.05$, closer to the Schwarzschild condition. Recall that the Schwarzschild and Ledoux conditions use different $(\partial S/\partial Y)$ derivatives, as shown in Figure~\ref{fig:eos_derivatives}. Consequently, the $R_\rho = 0.05$ model produces something of a ``helium ocean'' in the Saturn model (bottom left panel) and depletes more helium (bottom right panel), producing results similar to those found in \cite{Howard2024}. Moreover, only this model produces the classical result of helium rain atmospheric heating to explain Saturn's present-day luminosity, as demonstrated in Figure~\ref{fig:sat_lum_demo} \citep[e.g.,][]{FortneyHubbard2003, Mankovich2020}. As such, only the Schwarzschild-like convection criterion reproduces such atmospheric heating. By default, \orchard sets \texttt{r\_rho = 1.0} and \texttt{alpha\_rain = 0.1} (the equivalent of the yellow models in Figure~\ref{fig:jup_sat_he_rain}). Users can change this parameter, for example, to \texttt{r\_rho = 0.05}, and reproduce these demonstrations independently. We refer the reader to \cite{Sur2025b} for a detailed study and discussion of the $R_\rho$ parameter.

\begin{figure*}[!t]
\epsscale{1.05}
\plotone{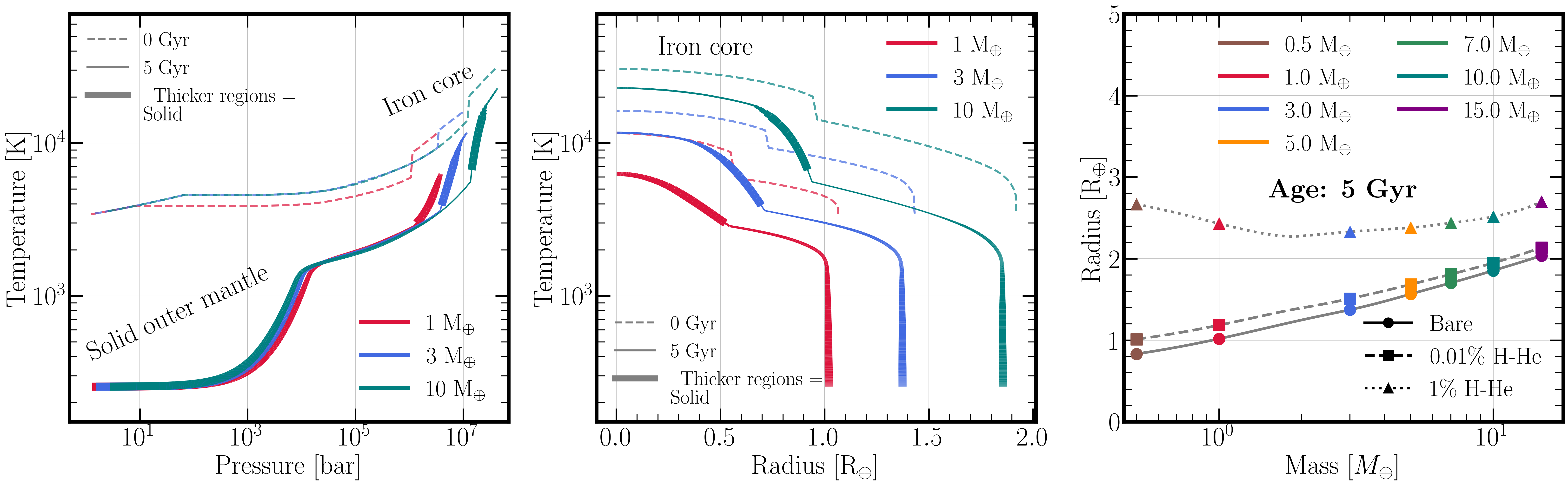}
\caption{Evolution demonstrations of 1, 3, and 10 \mearth rocky planets with no atmosphere (bare atmospheres), shown in the left two panels as a function of pressure and radial coordinate. The line thickness represents solidification (Equation~\ref{eq:melt_fraction}); thicker lines represent solid regions and thin lines liquid regions. The initial temperature profiles are plotted in thin dashed lines of the same color. As expected, smaller planets solidify their inner cores more readily than higher-mass planets, as indicated by the thick regions in the cores shown in the middle panel. The right panel compares the mass-radius (radius at 1 bar) relation of bare planets from 0.5 to 15 \mearth with 1) no envelope (solid segments), 2) 0.01\% by mass H-He envelopes (dashed segments), and 3) 1\% H-He by mass envelopes (dotted segments) at 5 Gyr.  An equilibrium temperature of 255 K was used for all models, yielding results similar to those in \cite{Lopez2014}. The mass/radius relation shows sensitivity to the amount of H-He in the envelope; smaller planets experience drastic radius enlargement even for 1\% H-He envelopes, a behavior seen in the top panel of Figure 1 of \cite{Lopez2014}.}
\label{fig:supearths_0.5_10}
\end{figure*}

Three rocky planet models of 1, 3, and 10 \mearth are showcased in Figure~\ref{fig:supearths_0.5_10}. A terrestrial iron core/mantle ratio of 1:2 is assumed for all super-Earths and sub-Neptune example calculations. The line thicknesses in the left and center panels, showing temperature vs. pressure and temperature vs. radius profiles, indicate the melt fraction; thicker lines indicate solidified regions. The rightmost panel shows an example mass-radius profile of these models. So-called ``bare'' models have no atmosphere, and they cool directly into space (see Section~\ref{subsec:atm}). The thin-envelope models modulate the cooling of each model, leading to significantly larger radii by 20-- 50\%, depending on mass, as shown in the right panel. This sensitivity to H-He fraction is also seen in Figure 1 of \cite{Lopez2014}. The mantle composition of these models is \olv, so the melt curve of \cite{Presnall1993} is used to assess mantle solidification via Equation~\ref{eq:melt_fraction}.  

\begin{figure}[!t]
\epsscale{1.0}
\plotone{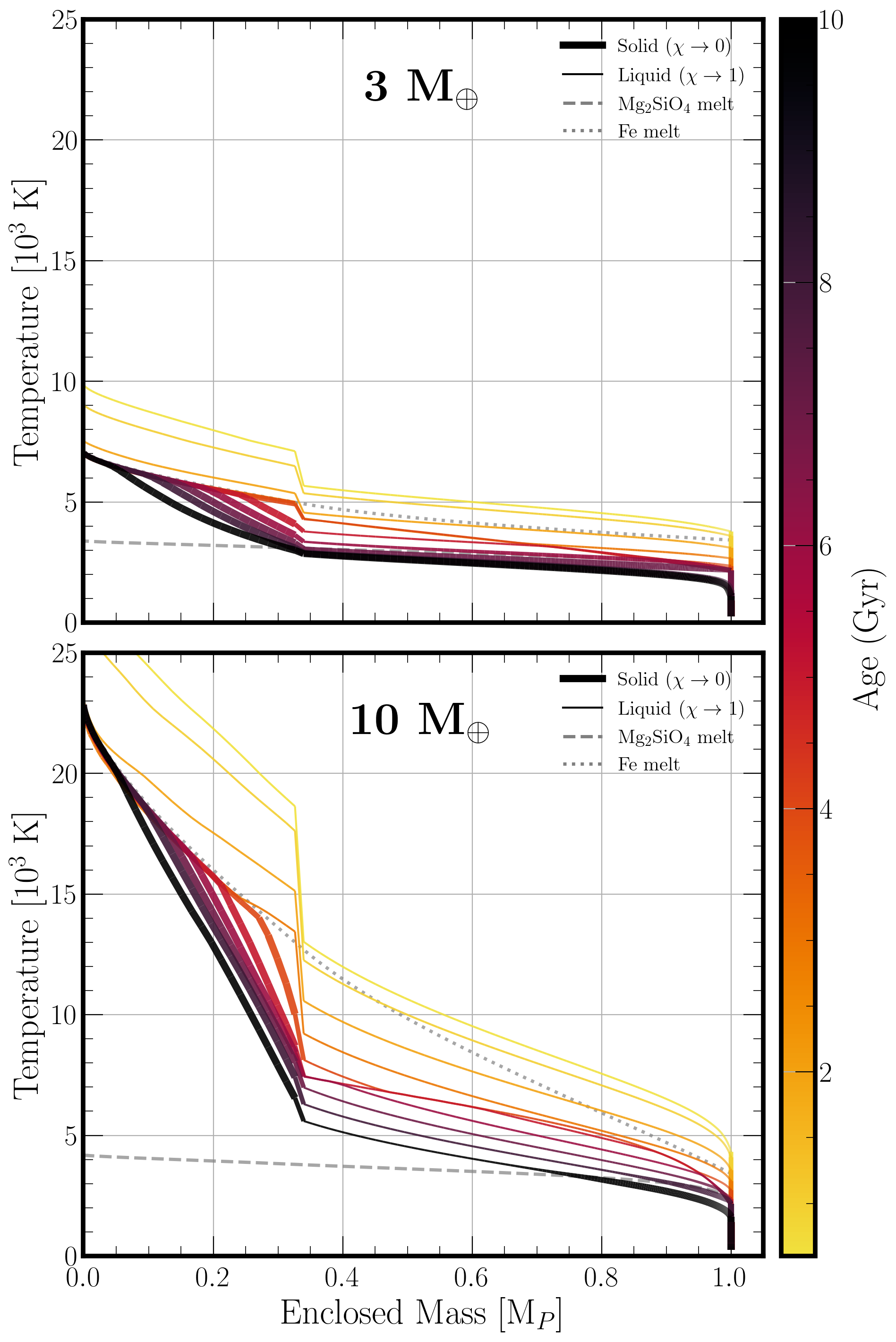}
\caption{Comparison of the thermal evolution history as a function of mass coordinate of a 3 (top) and 10 \mearth (bottom) model. The super-adiabatic temperature profiles are discontinuous at the core-mantle boundary since the mantle cools at a different rate than the core \citep{Tejada2026a}. Both harbor a thin, 0.01\% H-He by mass envelope or atmosphere. As with Figure~\ref{fig:mearth_melt_demo}, the faint dotted line is the melt curve of iron, the faint dashed line is the melt curve of \olv\, and each profile's line thickness indicates the degree of solidification, as indicated in each panel legend. Both models are initialized at the same mantle entropy, so they share a similar initial state. The iron melt curve is steeper at higher pressures, causing the 10 \mearth model to solidify its core sooner. On the other hand, the 3 \mearth model cools faster, so its mantle solidifies sooner. By 10 Gyr (black lines), the 3 \mearth model has almost completely solidified its mantle, but the 10 \mearth model has only solidified the upper 25\% of its mantle by mass.  }
\label{fig:3_5_melt_demo}
\end{figure}

Two super-Earth models of 3 and 10 \mearth are shown in Figure~\ref{fig:3_5_melt_demo}. The line thickness is a function of the melt fraction; thicker regions are more solid. The temperature jumps seen at early ages are due to inefficient cooling at the core-mantle boundary. These temperature jumps are expected in regions cooling at different rates, as discussed in \snI. We refer the reader to \snI\, for a more detailed description of these super-adiabatic profiles. Users have the option to start their structures with temperature continuity (\txt{fe\_core\_offset = 0}), but this will quickly produce a super-adiabatic temperature profile as the mantle cools faster than the core. In \txt{ORCHARD}, we ensure flux continuity, not temperature continuity, given that different regions can cool at different rates (See Figures 9 and 10 of \snI). The construction of these initial thermal states is described in Appendix~\ref{subsec:initial_setup}. The iron melt curves of \cite{Gonzalez-Cataldo2023} are shown as thin dotted lines and are described by Equation~\ref{eq:tmelt_fe}. This melt curve is steeper at higher pressures, so the core melting temperatures of the 10 \mearth model are $\sim$3 times that of the 3 \mearth model. The smaller 3\mearth is expected to cool faster and thus solidify its core sooner, but given the steepness of the melt temperatures at high pressures, these two models solidify their cores at approximately the same evolutionary age of 2.5 Gyr. Nevertheless, the mantle of the 3 \mearth model has completely solidified by 10 Gyr, while the mantle of the 10 \mearth model remains mostly liquid.

\begin{figure*}[!t]
\epsscale{1.05}
\plotone{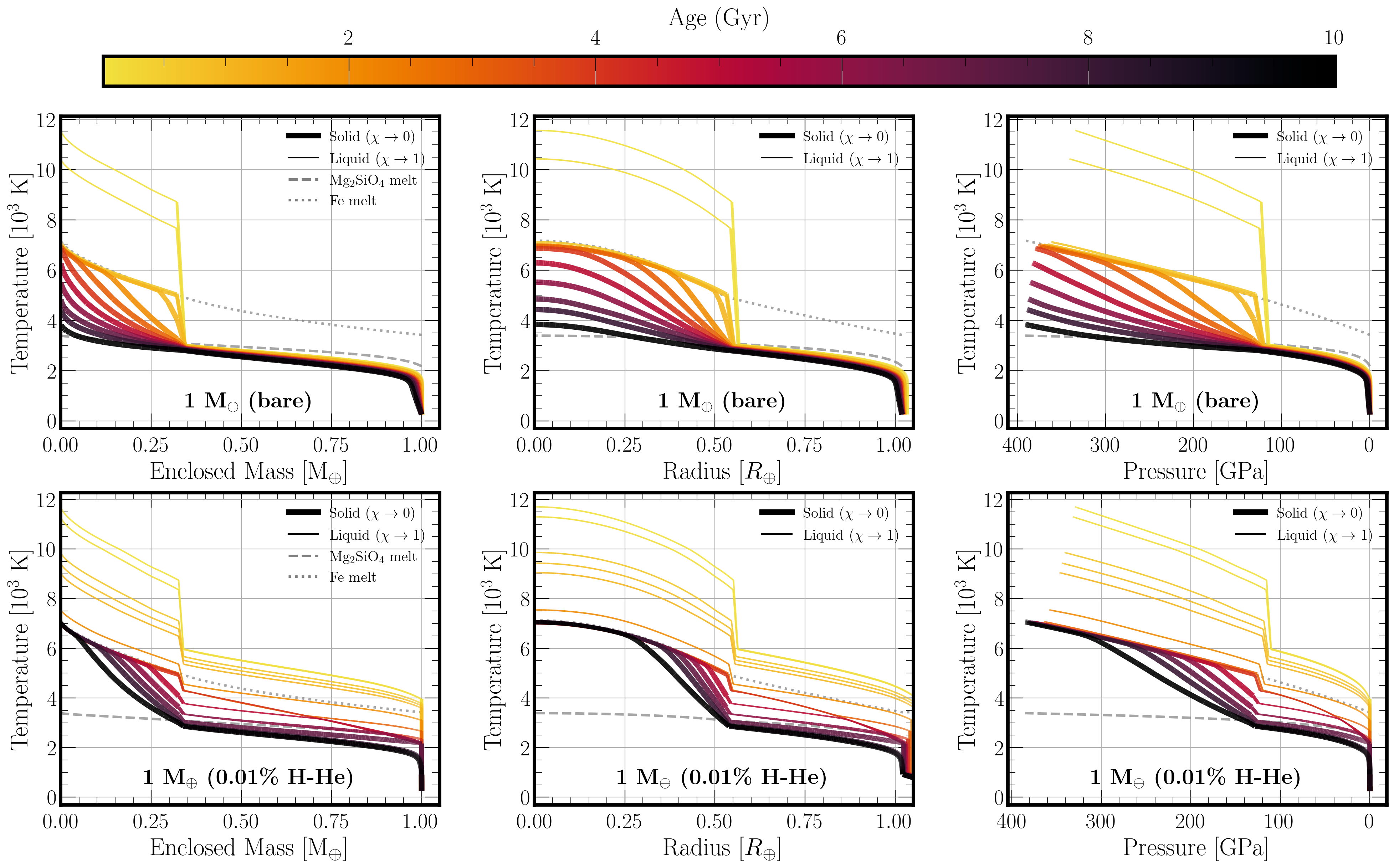}
\caption{Thermal evolution of a 1 \mearth model with no atmosphere (top row) and a model with a thin 0.01\% H-He by mass atmosphere/envelope. Thicker line segments indicate solidified regions, with thickness varying with the model's melt fraction, while thin line segments indicate liquid regions. The left, center, and right panels show the temperature evolution as a function of mass, radius, and pressure, respectively. The faint dotted line is the iron melt curve of \cite{Gonzalez-Cataldo2023}, and the faint dashed curve is the \olv\ melt curve of \cite{Presnall1993}. The mantles begin fully liquid and solidify within the first 1 Myr in the bare model, but the mantles of the thin atmosphere model begin solidifying at 3-4 Gyrs. The pure iron core follows the melt curve as it releases latent heat and begins to solidify. After the core fully solidifies, heat transport is solely by conduction, creating extended temperature gradients between the mantle and the inner, partially molten core. Inserting an envelope into the model, even if it comprises only 0.01\% of its mass, affects its thermal evolution by inhibiting cooling. }
\label{fig:mearth_melt_demo}
\end{figure*}

Two 1 \mearth models, with and without thin H-He rich atmospheres, are shown in the top and bottom rows of Figure~\ref{fig:mearth_melt_demo}, depicting the rates of core and mantle solidification. The temperature evolution is demonstrated with respect to the enclosed mass, radius, and pressure in each column. The thin H-He atmosphere in this model has only 0.01\% of the planet's total mass in H-He. Just as in Figures~\ref{fig:supearths_0.5_10} and \ref{fig:3_5_melt_demo}, thicker lines indicate solid regions, and thin line segments indicate liquid or partially melted regions.  The model without an atmosphere (bare; top row) solidifies its mantle almost instantly, so its mantle undergoes viscous convection (Section~\ref{subsec:visc_conv}) throughout its evolution, and its core begins to solidify by 2 Gyr. On the other hand, the thin H-He envelope model delays the onset of mantle solidification until $\sim$3 Gyr and does not experience full viscous convection until after 6 Gyr. By 6 Gyr, its core is fully solidified and no longer convects. The solidification of these models' cores generates large temperature gradients that are shallower than those at early ages. Upon solidification, the inner solid cores release latent heat and follow the melt curves (faint dotted lines) of \cite{Gonzalez-Cataldo2023}. Such characteristics of evolutionary models can be used to infer the lifetimes and strengths of terrestrial and super-Earth magnetic fields. 

\section{Areas for Future Improvement}\label{sec:improvements}

Many 1-D planetary evolution capabilities are yet to be included in a public version of \texttt{ORCHARD}. One of these, already applied in \snI, is silicate miscibility. They are not included here due to large uncertainties in the immiscibility of silicate and water. With improved testing, however, both silicate and water immiscibility might be fully deployed in later public versions of \texttt{ORCHARD}. In this section, we describe our current priorities for future evolutionary processes. 

\subsection{Rotational Effects on Convection}

Recently, \cite{Fuentes2023} showed that rotation can slow the convective velocity in gas giant planets by a factor of $\sim$6, leading to a reduction in kinetic energy of the convective fluid parcels by $\sim$200.
Consequently, the inclusion of rotational effects using the entrainment rates derived in \cite{Fuentes2023} can have an impact on the survivability of Jupiter and Saturn's fuzzy core interiors, and of those in gas giants generally. Such an effect has already been implemented in \texttt{MESA} and shown to slow convective velocities \citep{Agrawal2026}. Given its potential significance, incorporating adjusted mixing rates into the traditional MLT approach will be one goal for future realizations of \texttt{ORCHARD}. 

\subsection{Mass Loss}

Atmospheric and envelope mass loss is an important evolutionary process for low-density, highly irradiated sub-Neptune and hot Jupiter exoplanets \citep{Owen2013, Owen2017a, Gupta2019}. Mass loss due to a combination of photoevaporation and interior heat release \citep[e.g.,][]{Gupta2019, Tang2024b} has promise for explaining the radius valley among close-in super-Earth and sub-Neptune exoplanets \citep{Fulton2017}. Atmospheric mass loss prescriptions will be incorporated in future versions of \txt{ORCHARD} to facilitate hot sub-Neptune and hot Jupiter evolution studies.

\subsection{Core Erosion and Core Compositional Exchange}\label{subsec:comp_exchange}

Gas giant planets like Jupiter and Saturn are believed to have formed with large, compact central masses \citep{Bodenheimer1986, Pollack1996}. A possible cause of Jupiter's fuzzy core is the gradual erosion of this central core over its evolution \citep{Guillot2004}. However, this process has been found to be thermodynamically disfavored \citep[see Section 4.1 in][]{Helled2022b} and may thwart the mixing of the core \citep{Fuentes2025}. Nevertheless, the gradual erosion of the core needs to be incorporated into future versions of \orchard to enable more detailed studies of core erosion.

Beyond Jupiter and Saturn, compositional exchange between the cores and envelopes of sub-Neptunes is currently an active topic of research \citep[e.g.,][]{Nixon2025, Gupta2026}. Moreover, chemical reactions between the mantle and initial envelope compositions can affect the thermal and compositional evolution of sub-Neptune interiors \citep[e.g.,][]{Schlichting2022,Werlen2025a}. To properly model these exchanges, calculations of element equilibrium partitioning and an expanded suite of equations of state are required. Moreover, the associated chemical reaction rates need to be density-, pressure-, and temperature-dependent in order to inform the compositional state of the interior structure from one timestep to the next. Since these effects can likely affect the radii and atmospheric compositions of sub-Neptunes, incorporating this capability is another feature to include in future versions of \texttt{ORCHARD}. 

\subsection{Semiconvection Outlook}


Our internal tests of semi-convective implementations in fuzzy core/compositional gradient evolution have revealed the following challenges for preserving primordial fuzzy cores of gas giants: 1) Semiconvective regions, if they appear at all, appear early in the evolution. 2) When they appear in the early stages of evolution, they emerge and dissipate within $\sim$1-10 Myr timesteps, causing numerical instabilities. 3) When they persist longer than 100 Myr, they eventually become Ledoux-unstable due to cooling from above. Consequently, 4) they affect the evolution of compositional gradients at $\sim$ Gyr timescales only slightly. For now, our internal assessments agree with those presented in \cite{muller2020}, who also explored the challenge of preserving a primordial Jovian fuzzy core throughout its evolution and found that semiconvection has only modest effects on the evolution of fuzzy-core interiors.

\section{Conclusion}\label{sec:conclusion}

This paper introduces \texttt{ORCHARD}, a unified planetary evolution framework that synthesizes the methodologies of \apple (\papI) and \texttt{CMAPPER} (\seI), incorporating the ice giant capabilities introduced in \cite{Tejada2025b} and the rocky-planet thermodynamics/mixed-phase interior treatment of \cite{Tejada2026a}. The result is a publicly available code capable of modeling planets ranging from 0.5 \mearth to 10 \mjup, bridging the gap between existing gas-giant evolution codes \citep{Sur2024a, Helled2025b} and those designed for super-Earth-sized planets (\seI). Modeling the thermal and structural evolution of planets across the mass spectrum, from rocky super-Earths to massive gas giants, has never before been attempted using a single evolution code. 

The capabilities described in this work are designed to provide the exoplanet and planetary science communities with a common, flexible tool for modeling the interior evolution and static structures of planets across their respective scales of interest. Several of the most pressing open questions in planetary interiors, such as the evolution and possible emergence of fuzzy cores in Jupiter and Saturn, the likely bulk compositions of Uranus and Neptune, and the diverse radius and atmospheric observations of sub-Neptunes, should benefit from this modeling capability. By making \orchard publicly available, our aim is to lower the barrier to investigating broad questions in planetary evolution, to provide a general testbed for planet modeling, and to encourage a more integrated approach to planetary interior evolutionary efforts.

\section*{Data Availability}  

The data presented in this paper can be made available upon reasonable request to the authors. The full code, plus supporting files, will go live at \url{https://github.com/robtejada/orchard-public} upon acceptance of this paper. The \texttt{CMAPPER} super-Earth evolution code is available at \url{https://github.com/zhangjis/CMAPPER_rock}.

\begin{acknowledgments}
Special thanks are extended to Yi-Xian Chen for providing the atmospheric boundary conditions used in \orchard for gas-giant evolution, Kazumasa Ohno and Jonathan Fortney for providing their sub-Neptune/gas giant atmosphere models, Akash Gupta for lively discussions and critical feedback, Jisheng Zhang for developing and providing his evolution code \texttt{CMAPPER}, Yao Tang for valuable discussions on sub-Neptune interior evolution and mass loss processes, Mandy Bethkenhagen for providing her methane and ammonia EOSes, and Donghao Zheng and Jie Deng for providing \ppv EOS tables for \texttt{ORCHARD}, J.J. Dong for valuable discussions on the \ppv EOS, and Felipe Gonzalez-Cataldo for providing comprehensive solid and liquid iron EOSes. RTA further thanks Matthew Coleman at PICSciE for computational and technical assistance. This research was funded by the Center for Matter at Atomic Pressures (CMAP), a National Science Foundation (NSF) Physics Frontier Center under Award PHY-2020249. Any opinions, findings, conclusions, or recommendations expressed herein are those of the authors and do not necessarily reflect NSF views. Anthropic's Claude Code Opus 4.6, 4.8, and Fable 5 models were used to assist with code review, code documentation improvements, and internal debugging tests for \texttt{ORCHARD}. Agentic code applications were independently reviewed and scrutinized by the authors.
\end{acknowledgments}

\clearpage

\appendix

\section{Appendix: Numerical Methods}\label{sec:num_methods}

This section describes the implementation and discretization of the equations discussed in Sections~\ref{subsec:structure}, \ref{subsec:transport}, and \ref{subsec:comp_transport}. \orchard, like \apple, uses an operator-split approach in which the hydrostatic equilibrium and thermal and compositional flux updates are computed separately within the same timestep. Section~\ref{subsec:henyey} describes the Henyey relaxation method we deploy to calculate hydrostatic equilibrium (Equations~\ref{eq:1}---\ref{eq:2}). Then,  Section~\ref{subsec:transport_num} describes how we solve the energy and compositional transport equations (Equations~\ref{eq:3}---\ref{eq:6}). A detailed algorithmic flowchart of \orchard is included in Appendix B (Figure~\ref{fig:flowchart}), and a detailed algorithmic description, Algorithm~\ref{alg:evolution}, is provided in the same Appendix. 

\subsection{Henyey Relaxation Method}\label{subsec:henyey}

Stellar evolution codes have long relied on the
Henyey relaxation scheme \citep{Henyey1964}, which solves the full set of structural equations simultaneously on a one-dimensional grid. We refer the reader to Chapter 12.2 of \cite{Kippenhahn2012} for a general overview.\footnote{We also refer to: \url{https://www.astro.sunysb.edu/lattimer/PHY521/henot.pdf}.}  The Henyey Newton-Raphson iteration linearizes the residual of the discretized equations and inverts the resulting block-tridiagonal Jacobian, achieving
quadratic convergence near the solution.

The 1-D planetary structure in \orchard is defined from the outside-in and is divided into three structural regions indexed by mass coordinate $m$, denoted in \orchard by \texttt{kcore} and \texttt{kcore\_fe}. The mass grid index $k_{\rm core}$ is the envelope-mantle boundary, and $k_{\rm core, Fe}$ is the core-mantle boundary. Users can specify coreless gas giants ($k_{\rm core}=N$), planets
with no iron core ($k_{\rm core,Fe}=N$), and bare-rock super-Earths ($k_{\rm core}=0$), where $N$ is the number of mass shells. The latter two cases are automatically defined, respectively, by setting the iron core mass to 0 or setting the total mass equal to the core mass.  This layered assumption will be relaxed in the future to allow compositional exchange between the mantle and the envelope (see Section~\ref{subsec:comp_exchange}).

For $N$ mass shells, the state vector $\mathbf{Y}$ has $2N$ components,
\begin{equation}\label{eq:statevec}
  \begin{split}
    \mathbf{Y} = \bigl(&\ln P_0,\;\ln r_0,\;\ln P_1,\;\ln r_1, \\
                       &\;\ldots,\;\ln P_k,\;\ln r_k,\;\ldots,\ln P_{N-1},\;\ln r_{N-1}\bigr)^{\!\top},
  \end{split}
\end{equation}
where $P_k$ is the pressure at the center of zone~$k$ and $r_k$ is the radius of its outer boundary.  Even-indexed elements carry $\ln P$ and odd-indexed elements carry $\ln r$. The thermodynamic quantities that close the system with temperature $T_k(S_k, P_k, Y_k, Z_k)$ and density $\rho_k(S_k, P_k, Y_k, Z_k)$ are evaluated with the EOS at each iteration and are treated as auxiliary variables updated from the current $(S, P, Y, Z)$ state.

The residual vector $\mathbf{A}$ contains $2N$ equations; one pressure (hydrostatic equilibrium) equation and one radius (mass continuity) equation per zone.  We define auxiliary logarithmic variables
\begin{equation}
  x_k \equiv \ln r_k, \qquad
  y_k \equiv \ln P_k, \qquad
  q_k \equiv \ln \rho_k\, .
\end{equation}
The mass grid boundaries $m_k$ ($k = 0,\ldots,N$) are fixed input with $m_0 = M_{\rm planet}$ at the surface and $m_N = 0$ at the center. 

For each interior zone, the pressure difference between adjacent shells balances gravity and centrifugal force, while the radius difference is just the mass contained in that given shell (see Equations~\ref{eq:1} and \ref{eq:2}). The associated difference equations are:
\begin{align}\label{eq:Ak_interior}
  A_{2k} &= y_{k-1} - y_k
             \notag\\
         &\quad
           + \frac{G}{4\pi}\,\frac{\Delta m_k^{\star}}{2}\,m_k\,
             e^{-\frac{1}{2}(y_k + y_{k-1}) - 4x_k}
           \notag\\
         &\quad
           - \frac{\Delta m_k^{\star}}{2}\,\frac{\omega_k^2}{6\pi}\,
             e^{-x_k - \frac{1}{2}(y_{k-1}+y_k)},\\
A_{2k+1} &= x_{k} - x_{k+1} 
           \notag\\
         &\quad- \frac{(m_{k} - m_{k+1})}{4 \pi} e^{-q_{k} - \frac{3}{2}(x_k + x_{k+1})}\, ,
\end{align}
where $\Delta m_k^{\star} \equiv m_{k-1} - m_{k+1}$ spans two shell boundaries, as done in \mesa \citep{Paxton2011, Paxton2013, Paxton2018}. The Newton correction $\delta\mathbf{Y}$ at each iteration satisfies
\begin{equation}\label{eq:newton}
  \mathbf{J}\,\delta\mathbf{Y} = \mathbf{A}\, ,
\end{equation}
where $\mathbf{J} = \pp\mathbf{A}/\pp\mathbf{Y}$ is the $2N\times 2N$ Jacobian matrix.  Because each residual equation couples at most three adjacent zones, $\mathbf{J}$ has a sparse, block-tridiagonal structure with $2\times 2$ blocks.

For an interior zone $k\in\{1,\ldots,N-2\}$, the non-zero Jacobian entries of the pressure equation ($i = 2k$) are:
\begin{align}
  J_{i,\,i-2}
  &= \frac{\pp A_{2k}}{\pp y_{k-1}}
   = 1 - \frac{G \Delta m_k^{\star}\,m_k\,}{16\pi}\,
             e^{-\frac{1}{2}(y_k + y_{k-1}) - 4x_k}
             \notag\\
             &\quad + \frac{\omega_k^2\,\Delta m_k^{\star}}{24\pi}\,
     e^{-x_k - \frac{1}{2}(y_{k-1}+y_k)},
  \label{eq:J_pk_left}
\\[4pt]
  J_{i,\,i}
  &= \frac{\pp A_{2k}}{\pp y_k}
   = -1 - \frac{G\,\Delta m_k^{\star}\,m_k}{16\pi}\,
     e^{-\frac{1}{2}(y_k + y_{k-1}) - 4x_k}
  \notag\\
  &\quad + \frac{\omega_k^2\,\Delta m_k^{\star}}{24\pi}\,
     e^{-x_k - \frac{1}{2}(y_{k-1}+y_k)},
  \label{eq:J_pk_self}
\end{align}

\begin{align}
  J_{i,\,i+1}
  &= \frac{\pp A_{2k}}{\pp x_k}
   = -\frac{G\,\Delta m_k^{\star}\,m_k}{2\pi}\,
     e^{-\frac{1}{2}(y_k + y_{k-1})- 4x_k}
  \notag\\
  &\quad + \frac{\omega_k^2\,\Delta m_k^{\star}}{12\pi}\,
     e^{-x_k - \frac{1}{2}(y_{k-1}+y_k)}\, .
  \label{eq:J_pk_right}
\end{align}
Here, $J_{i,\,i-2}$ couples to $\ln P_{k-1}$ (two state-vector
positions to the left), $J_{i,\,i}$ is the diagonal ($\ln P_k$), and $J_{i,\,i+1}$ couples to $\ln r_k$ (one position to the right).

The radius equation ($i = 2k+1$) has two non-zero entries:

\begin{align}
  J_{i,\,i}
  &= \frac{\pp A_{2k+1}}{\pp x_k}
   = 1 + \frac{3(m_k - m_{k+1})}{8\pi}\,
     e^{-\frac{3}{2}(x_k+x_{k+1}) - q_k},
  \label{eq:J_rk_self}
\\[4pt]
  J_{i,\,i+2}
  &= \frac{\pp A_{2k+1}}{\pp x_{k+1}}
   = -1 + \frac{3(m_k - m_{k+1})}{8\pi}\,
     e^{-\frac{3}{2}(x_k+x_{k+1}) - q_k}.
  \label{eq:J_rk_right}\, .
\end{align}
The surface and innermost boundaries are computed separately, and we refer the reader to Equations B5 through B8 of \papI\, for reference. We then solve Equation~\ref{eq:newton} via matrix inversion to update the state vector, $\mathbf{Y}^{n+1} = \mathbf{Y}^n - \delta\mathbf{Y}$. 

\subsection{Heat Transport Updates}\label{subsec:transport_num}

All quantities entering the Newton solver are cast in dimensionless form to improve the conditioning of the $3N\times 3N$ linear system. Four reference scales are introduced:
\begin{align}
  S_0 &= \kbbar
       &&\text{(entropy scale)},
  \label{eq:S0}\\
  T_0 &= 1000\;\mathrm{K}
       &&\text{(temperature scale)},
  \label{eq:T0}\\
  M_0 &= M_{\rm planet}
       &&\text{(mass scale)},
  \label{eq:M0}\\
  R_0 &= R_{\rm planet}
       &&\text{(radius scale)}\, .
  \label{eq:R0}
\end{align}
$M_0$ and $R_0$ are set to the planet's total mass and initial radius (e.g.\ $M_{\rm Jup}$ and $R_{\rm Jup}$ for a Jupiter model). The dimensionless state variables are
\begin{equation}\label{eq:dimless_vars}
  S' = \frac{S}{S_0},\,
  T' = \frac{T}{T_0},
  \dd m' = \frac{\dd m}{M_0},
  \dd r'_i = \frac{r_{k+1} - r_k}{R_0}\, ,
\end{equation}
where $S_0$ is a normalization constant.  $Y$ and $Z$ are already dimensionless mass fractions and are included in the state vector without rescaling.

The state vector has $3N$ components, interleaving entropy,
helium, and metallicity for each cell:
\begin{equation}\label{eq:state_vec}
  \mathbf{Z} = (S'_0, Y_0, Z_0,\; S'_1, Y_1, Z_1,\; \ldots,\;
                S'_{N-1}, Y_{N-1}, Z_{N-1})\, .
\end{equation}
The residual vector $\mathbf{B}(\mathbf{Z})$ is assembled from the implicit (backward-Euler) discretization of
Equations~(\ref{eq:entropy_master})--(\ref{eq:conv_mixing}).

For brevity, we show here only the discretization of the entropy equation:

\begin{align}
  B_{3k}   &= (S'_k - S'_{k,\rm old})
    + \frac{\Delta t}{\dd m'_k\,T'_k}\,\frac{\Delta F_k}{M_0T_0S_0}\\
        &\qquad\quad
    + \frac{\sum_i(\pp U/\pp X_i)\,(X_{i,k} - X_{i,k,\rm old})}{T'_k\,T_0 S_0} \\        
       &\qquad\quad
    - \frac{\Delta t}{\dd m'_k\,T'_k}\,H'_k + \frac{\mathcal{L}}{T_0\,S_0}\,\frac{\Delta\chi_k}{T'_k}\, ,
    \label{eq:BS}\\
\end{align}
where $\Delta\chi_k = \chi_k^{n+1} - \chi_k^n$ is the melt fraction change, $\Delta F_k = F_{k+1} - F_k$ is the net flux divergence across cell~$k$, and $H_k'$ is the radiogenic heat (Equation~\ref{eq:Hspec}).

The coupled $3N\times 3N$ system $\mathbf{B}(\mathbf{Z})=\mathbf{0}$ is solved by the Newton-Raphson solution of the next timestep's state vector, $\mathbf{z}^{n+1}$:
\begin{equation}\label{eq:NR}
  \mathbf{J}^{n}\,\delta\mathbf{Z}
  = \mathbf{B}^n,
  \qquad
  \mathbf{Z}^{n+1}
  = \mathbf{Z}^n - \delta\mathbf{Z}\, .
\end{equation}

\clearpage

\section{Appendix: Code Description}\label{sec:code_description}

All of the parameter files for \orchard are \txt{.ini} files, or initialization/configuration files. The main configuration file is called \href{    https://github.com/robtejada/orchard/blob/main/parameters_default.ini}{\texttt{parameters\_default.ini}}, which contains all available user parameters, and the descriptions are in \href{https://github.com/robtejada/orchard/blob/main/parameter_descriptions.md}{\texttt{parameter\_descriptions.md}}. User example parameter files to evolve super-Earth, sub-Neptunes (rocky or water world), gas giants, and Solar System planets are found in \href{https://github.com/robtejada/orchard/tree/main/parameter_examples}{\texttt{parameter\_examples}} in the home \href{https://github.com/robtejada/orchard}{\texttt{orchard}} repository. \orchard can be run easily in your command line terminal. 

\subsection{Example user parameter configuration file}\label{sec:parameter_config}

The default user parameter provided in the GitHub repository can be run as follows:

\begin{lstlisting}[language=bash]
python evolution.py --config parameters_user.ini
\end{lstlisting}
where \texttt{parameters\_user.ini} is a basic, homogeneous and adiabatic 1 M$_J$ model provided as the first example evolution run. Its parameter configuration file is:

\begin{lstlisting}[language=bash]
# ORCHARD user-facing parameter file
#
# This file is read after `parameters/parameters_default.ini` (via `--config`),
# so any parameter not listed here automatically keeps the default value.
#
# Usage:
#   python evolution.py --config parameters/parameter_user.ini
#
# Edit only the options below for typical runs, such as the total mass (M_MJup). Advanced controls remain in
# `parameters_default.ini`.

[general]
# Grid resolution (higher = slower, usually more accurate)
N = 500
# End time of the simulation [Gyr]
final_age = 10.0
# Save interval [Myr]; 0 saves every accepted step
save_interval = 0
# Maximum timestep [Myr]
max_step = 50
# Optional HDF5 output (text outputs are still used by many utilities)
save_hdf5 = True
hdf5_filename = evolution.h5

[equation_of_state]
# H-He EOS: cms, cd, scvh
hhe_eos = cd
# Envelope heavy-element EOS (default shown)
z_eos = aqua

[initial]
# Change total mass here (in Jupiter masses)
M_MJup = 1.0
# Initial entropy [k_B / baryon]
S_ini = 10.0
# Initial helium fraction Y / (X + Y)
Y_ini = 0.277
# Initial envelope heavy element mass fraction Z (default 3.16 solar shown)
Z_ini = 0.05181726985106876
\end{lstlisting}

A terrestrial, 1\mearth example model parameter requires different parameters than a default, 1.0 \mjup model, such as a non-isothermal compact core (i.e., our whole planet in this case), and it should include radioactive and latent heat source terms (\texttt{radioactive = True}). A typical configuration file can be

\begin{lstlisting}[language=bash]
[general]
# Grid resolution
N = 500
# place more zones near the surface for better resolution
surf_width = 8e-2
surf_width2 = 6.5e-3
# Maximum timestep [Myr]
max_step = 50
# Print detailed per-step diagnostics (errors, iterations, energy metrics)
verbose = True
save_verbose = True
# Optional HDF5 output (text outputs are still used by many utilities)
save_hdf5 = True
hdf5_filename = evolution.h5

[hydrostatic_equilibrium]
# If False, core/mantle entropy is set from [core] values instead of matching envelope
env_s_start = False
isothermal_compact_core = False

[boundary_condition]
# Examples: Jupiter, Saturn, Uranus, Neptune, Sub_Neptune, Super_Earth, Super_Jupiter
planet = Super_Earth
# Atmosphere BC model (depends on planet): c23, c26, f11, f07, g75, gray, bare
bc_atm = bare
T_eq = 400
bond_albedo = 0.3

[initial]
# Planet mass in Earth masses
M_Mearth = 1.0

[transport]
radioactive = True
latent_heat_effect = True

[core]
# Core masses [Earth masses]-- coreless default
# Core composition / thermal state (advanced users may tune these)
mantle_entropy = 0.51
# Same as total mass above
mass_core = 1.0
# Mass of iron core [Earth masses]
mass_core_fe = 0.333
# Iron core entropy offset in k_B/baryon. Sets hotter initial iron core if desired.
# Select offset of 0.0 for no initial jump.
fe_core_offset = 0.035
# Core/mantle material model choices
mantle_comp = mg2sio4
eos_mantle = comb
core_comp = Fe_pure
# Equation of state of iron core (Ichikawa et al. 2014)
eos_core = D17_comb
viscous_mantle_convection = True
mantle_thermal_conductivity = None
# Conductivity of the core. This is a constant value for now, but can be made T and rho dependent in the future.
core_thermal_conductivity = 40
\end{lstlisting}

We recommend the user read the installation instructions and the parameter descriptions. To learn to use the code for your own purposes, please see the list of \href{https://github.com/robtejada/orchard/tree/main/tutorials}{\texttt{tutorials}}.

\subsection{\orchard code structure}

This section describes the main evolution files in the \orchard GitHub repository. All of these files are described in detail in the repository. For clarity, we also provide detailed documentation in each file. The following \txt{Python} scripts comprise the core structure of \orchard:

\begin{itemize}
    \item \txt{initial.py}: Builds the specified initial thermal and compositional structure, initializes the EOSes described in \ref{subsec:EOS}, and builds the mass grid. A detailed description of the initialization workflow is provided in Appendix~\ref{subsec:initial_setup}.
    \item \txt{atm\_bc.py}: Implements the atmosphere boundary conditions to provide $\Tint$, $\Teff$, and their associated derivatives as described in \ref{subsec:atm}.
    \item \txt{rk4\_initial.py}: Initializes the user parameter model with a Runge-Kutta integrator to fourth order, improving initial hydrostatic equilibrium steps for higher mass or rocky planet models. The user needs to set \txt{rk4\_initial = True} to use it. If \txt{rk4\_initial = False}, then the initial structure is a 9 $ k_B$ baryon$^{-1}$ isentrope that works for most gas-giant planet models.
    \item \txt{hydrostatic.py}: Updates the radius, pressure, density, and temperature from the current compositional/entropy state using Equations~\ref{eq:1}--\ref{eq:5} with a Henyey relaxation solver, incorporating the optional rotational (\txt{rotation = True}) terms using theory of figures to fourth order using ToF4 \citep{Nettelmann2017}, found in \txt{TOF.py}. We set a default matrix inversion convergence tolerance of 10$^{-10}$ (\txt{tol\_hydro = 1e-10}) in the hydrostatic solver, which is editable via the user configuration file.
    \item \txt{transport.py}: Contains the global Newton-Raphson matrix solver that advances the entropy and composition ($S$, $Y$, $Z$) by computing the convective, radiative, conductive, and compositional fluxes as described in Section~\ref{subsec:transport}, with optional miscibility of helium or metals (\txt{he\_rain = True}, \txt{metal\_rain = True}), and calculates the surface losses, all implicitly. We set a default matrix inversion convergence tolerance of 10$^{-10}$ (\txt{tol\_NR = 1e-10}) for the Newton-Raphson matrix solver, which is editable via the user configuration file.
    \item \txt{evolution.py}: This is the top-level driver that updates the thermodynamic and structural variables ($S$, $Y$, $Z$, $T$, $\rho$, $P$, $R$, etc.). This module executes the operator-split evolution loop with \txt{transport.update\_SYZ(...)}, then \txt{hydrostatic.hydrostatic\_equilibrium(...)}), and manages adaptive timesteps according the relative errors of $S$, $Y$, $Z$, $T$, $\rho$, $\chi$ (melt fraction). Finally, this module saves the evolution data in either \txt{.txt} or \txt{.h5} file format, set by the user (\txt{save\_hdf5 = True}). 
\end{itemize}

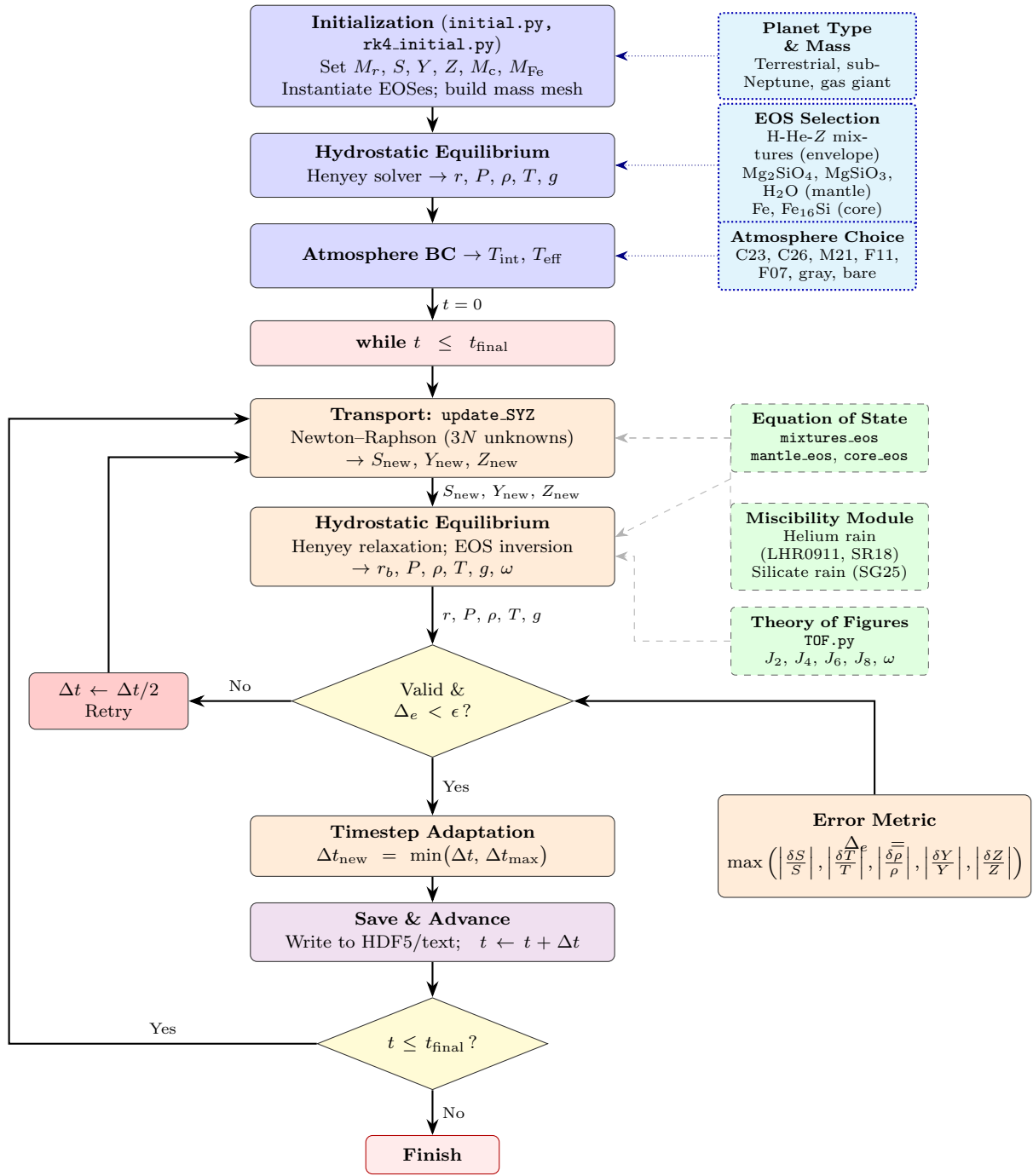
\begin{figure*}[!t]
\centering
\begin{tikzpicture}[
    init/.style={rectangle, draw=black!80, fill=blue!15, text width=5.4cm,
                 minimum height=1.0cm, align=center, rounded corners=3pt,
                 font=\footnotesize},
    process/.style={rectangle, draw=black!80, fill=orange!15, text width=5.4cm,
                    minimum height=0.9cm, align=center, rounded corners=3pt,
                    font=\footnotesize},
    decision/.style={diamond, draw=black!80, fill=yellow!20, text width=2.0cm,
                     minimum height=0.6cm, align=center, aspect=2.5,
                     font=\footnotesize},
    auxiliary/.style={rectangle, draw=black!60, fill=green!12, text width=2.8cm,
                      minimum height=0.6cm, align=center, rounded corners=2pt,
                      font=\scriptsize, dashed},
    userbox/.style={rectangle, draw=blue!70!black, fill=cyan!10, text width=2.8cm,
                    minimum height=0.6cm, align=center, rounded corners=2pt,
                    font=\scriptsize, densely dotted, thick},
    loopbox/.style={rectangle, draw=black!80, fill=red!10, text width=5.4cm,
                    minimum height=0.7cm, align=center, rounded corners=3pt,
                    font=\footnotesize\bfseries},
    retrybox/.style={rectangle, draw=black!80, fill=red!20, text width=2.2cm,
                     minimum height=0.7cm, align=center, rounded corners=3pt,
                     font=\footnotesize},
    savebox/.style={rectangle, draw=black!80, fill=violet!12, text width=5.4cm,
                    minimum height=0.8cm, align=center, rounded corners=3pt,
                    font=\footnotesize},
    finishbox/.style={rectangle, draw=red!70!black, fill=red!8, text width=1.8cm,
                      minimum height=0.6cm, align=center, rounded corners=3pt,
                      font=\footnotesize\bfseries},
    arrow/.style={-{Stealth[length=2.5mm]}, thick},
    dasharrow/.style={-{Stealth[length=2mm]}, dashed, gray!60},
    dotarrow/.style={-{Stealth[length=2mm]}, densely dotted, blue!50!black},
    every node/.append style={inner sep=4pt},
]


\node[init] (init) {
    \textbf{Initialization} (\texttt{initial.py, rk4\_initial.py})\\[1pt]
    Set $M_r$, $S$, $Y$, $Z$, $M_{\rm c}$, $M_{\rm Fe}$\\
    Instantiate EOSes; build mass mesh
};

\node[init, below=0.4cm of init] (inithse) {
    \textbf{Hydrostatic Equilibrium}\\[1pt]
    Henyey solver $\rightarrow$ $r$, $P$, $\rho$, $T$, $g$
};

\node[init, below=0.4cm of inithse] (initbc) {
    \textbf{Atmosphere BC} $\rightarrow$ $T_{\rm int}$, $T_{\rm eff}$
};


\node[userbox, right=1.6cm of init] (planettype) {
    \textbf{Planet Type \& Mass}\\
    Terrestrial, sub-Neptune, gas giant
};

\node[userbox, right=1.6cm of inithse] (eoschoice) {
    \textbf{EOS Selection}\\
    H-He-$Z$ mixtures (envelope)\\
    Mg$_2$SiO$_4$, MgSiO$_3$, H$_2$O (mantle)\\
    Fe, Fe$_{16}$Si (core)
};

\node[userbox, right=1.6cm of initbc] (bcchoice) {
    \textbf{Atmosphere Choice}\\
    C23, C26, M21, F11,\\
    F07, gray, bare
};


\node[loopbox, below=0.5cm of initbc] (loopstart) {
    \textbf{while} $t \leq t_{\rm final}$
};

\node[process, below=0.5cm of loopstart] (transport) {
    \textbf{Transport:} \texttt{update\_SYZ}\\[1pt]
    Newton--Raphson ($3N$ unknowns)\\
    $\rightarrow$ $S_{\rm new}$, $Y_{\rm new}$, $Z_{\rm new}$
};

\node[process, below=0.45cm of transport] (hydro) {
    \textbf{Hydrostatic Equilibrium}\\[1pt]
    Henyey relaxation; EOS inversion\\
    $\rightarrow$ $r_b$, $P$, $\rho$, $T$, $g$, $\omega$
};

\node[decision, below=0.9cm of hydro] (errcheck) {
    Valid \&\\$\Delta_e < \epsilon$\,?
};

\node[retrybox, left=1.6cm of errcheck] (retry) {
    $\Delta t \leftarrow \Delta t / 2$\\
    Retry
};

\node[process, below=0.9cm of errcheck] (dtadapt) {
    \textbf{Timestep Adaptation}\\[1pt]
    $\Delta t_{\rm new} = \min\!\big( \Delta t,\,\Delta t_{\rm max}\big)$
};

\node[process, right=1.6cm of dtadapt, text width=4.6cm, minimum height=1.6cm] (errcomp) {
    \textbf{Error Metric}\\[1pt]
    $\Delta_e = $\\[-3pt]
    $\max\left(\left|\frac{\delta S}{S}\right|,
    \left|\frac{\delta T}{T}\right|,
    \left|\frac{\delta \rho}{\rho}\right|,
    \left|\frac{\delta Y}{Y}\right|,
    \left|\frac{\delta Z}{Z}\right|\right)$
};

\node[savebox, below=0.4cm of dtadapt] (save) {
    \textbf{Save \& Advance}\\[1pt]
    Write to HDF5/text;\quad $t \leftarrow t + \Delta t$
};

\node[decision, below=0.55cm of save] (timecheck) {
    $t \leq t_{\rm final}$\,?
};

\node[finishbox, below=0.7cm of timecheck] (finish) {
    \textbf{Finish}
};


\node[auxiliary, right=1.8cm of transport] (eos) {
    \textbf{Equation of State}\\
    \texttt{mixtures\_eos}\\
    \texttt{mantle\_eos}, \texttt{core\_eos}
};

\node[auxiliary, right=1.8cm of hydro] (misc) {
    \textbf{Miscibility Module}\\
    Helium rain (LHR0911, SR18)\\
    Silicate rain (SG25)
};

\node[auxiliary, below=0.25cm of misc] (tof) {
    \textbf{Theory of Figures}\\
    \texttt{TOF.py}\\
    $J_2$, $J_4$, $J_6$, $J_8$, $\omega$
};


\draw[arrow] (init) -- (inithse);
\draw[arrow] (inithse) -- (initbc);
\draw[arrow] (initbc) -- node[right, font=\scriptsize] {$t=0$} (loopstart);
\draw[arrow] (loopstart) -- (transport);
\draw[arrow] (transport) -- node[right, font=\scriptsize] {$S_{\rm new}$, $Y_{\rm new}$, $Z_{\rm new}$} (hydro);
\draw[arrow] (hydro) -- node[right, font=\scriptsize] {$r$, $P$, $\rho$, $T$, $g$} (errcheck);
\draw[arrow] (errcheck) -- node[right, font=\scriptsize] {Yes} (dtadapt);
\draw[arrow] (dtadapt) -- (save);
\draw[arrow] (save) -- (timecheck);
\draw[arrow] (timecheck) -- node[right, font=\scriptsize] {No} (finish);

\draw[arrow] (errcheck.west) -- node[above, font=\scriptsize] {No} (retry);
\draw[arrow] (retry.north) |- ([yshift=-0.3cm]transport.west);

\draw[arrow] (timecheck.west) -- node[above, font=\scriptsize] {Yes}
    ++(-4.8,0) |- ([yshift=0.3cm]transport.west);

\draw[arrow] (errcomp.north) |- (errcheck.east);


\draw[dasharrow] (eos.west) -- (transport.east);
\draw[dasharrow] (misc.west) |- (transport.east);
\draw[dasharrow] ([yshift=-0.1cm]eos.south west) -- ([yshift=0.15cm]hydro.east);
\draw[dasharrow] (tof.west) -| ([xshift=0.3cm, yshift=-0.15cm]hydro.east) -- ([yshift=-0.15cm]hydro.east);


\draw[dotarrow] (planettype.west) -- (init.east);
\draw[dotarrow] (eoschoice.west) -- (inithse.east);
\draw[dotarrow] (bcchoice.west) -- (initbc.east);

\end{tikzpicture}
\caption{Flowchart of the \textsc{orchard} evolution code, adapted and updated from \papI. Blue boxes show the initialization phase; orange boxes denote the main computational steps within each timestep; yellow diamonds are decision points for step acceptance and loop termination; the violet box marks data output, and green dashed boxes indicate auxiliary modules called by the core routines. If the combined validity and error check fails (red path), the timestep is halved, and the step is retried. The error metric $\Delta_e$ is the maximum fractional change in entropy, temperature, density, helium fraction, and metal fraction, computed over zones above a configurable pressure floor; validity requires convergence of the hydrostatic equilibrium solve with no NaN or non-monotonic pressure.}
\label{fig:flowchart}
\end{figure*}

We provide a few auxiliary modules that set user-defined initial $S$, $Y$, and $Z$ profiles (\texttt{profiles.py}) and define mantle properties, including electrical and thermal conductivities and viscosities (\texttt{mantle\_core\_props.py}). These and additional auxiliary files are housed in the \texttt{utils} folder within the \texttt{orchard} folder. We now provide a detailed graphic and algorithmic workflow in Figure~\ref{fig:flowchart}. This is a recapitulation of Algorithm~\ref{alg:evolution}. 

\begin{algorithm*}[ht!]
\DontPrintSemicolon
\SetAlgoLined
\SetKwInput{KwInput}{Input}
\SetKwInput{KwOutput}{Output}
\SetKwFunction{InitGrid}{InitializeGrid}
\SetKwFunction{HSE}{HydrostaticEquilibrium}
\SetKwFunction{UpdateSYZ}{UpdateSYZ}
\SetKwFunction{AtmBC}{AtmosphereBC}

\KwInput{$M_p$, $S_0$, $Y_0$, $Z_0$, $M_{\rm c}$, $M_{\rm Fe}$, $t_{\rm final}$, $\Delta t_0$, $\epsilon$}
\KwOutput{Profiles $r(m,t)$, $P(m,t)$, $\rho(m,t)$, $T(m,t)$, $S(m,t)$, $Y(m,t)$, $Z(m,t)$}

\BlankLine
\tcc{Initialization: build mass grid, composition, EOS, and initial structure}
$(r_b,\, P,\, m_b,\, S,\, Y,\, Z,\, k_{\rm c},\, k_{\rm Fe}) \leftarrow$ \InitGrid{$N,\, M_p,\, S_0,\, Y_0,\, Z_0$}\;
$(r_b,\, P,\, \rho,\, T,\, g) \leftarrow$ \HSE{$S,\, P,\, m_b,\, Y,\, Z$}\;
$(T_{\rm int},\, T_{\rm eff}) \leftarrow$ \AtmBC{$S,\, Y,\, Z,\, g$}\;
$t \leftarrow 0$;\quad $\Delta t \leftarrow \Delta t_0$\;

\BlankLine
\tcc{Main evolution loop with adaptive timestepping}
\While{$t \leq t_{\rm final}$}{
    \tcc{Transport: global Newton--Raphson for $S$, $Y$, $Z$ ($3N$ unknowns)}
    $(S',\, Y',\, Z') \leftarrow$ \UpdateSYZ{$T,\, P,\, \rho,\, m_b,\, \Delta t,\, S,\, Y,\, Z$}\;
    \quad includes convective, radiative, conductive, and semiconvective fluxes, plus He/Z rain\;

    \BlankLine
    \tcc{Hydrostatic equilibrium: Henyey relaxation with EOS inversion}
    $(r_b',\, P',\, \rho',\, T',\, g) \leftarrow$ \HSE{$S',\, P,\, m_b,\, Y',\, Z'$}\;

    \BlankLine
    \tcc{Error control: maximum fractional change over pressure-masked interior}
    $\Delta_e =\max\left(|\delta S/S|,\; |\delta Y/Y|,\; |\delta Z/Z|,\; |\delta T/T|,\; |\delta\rho/\rho|\right)$\;

    \eIf{\textnormal{state invalid (NaN, non-monotonic $P$) or } $\Delta_e > \epsilon/100$}{
        $\Delta t \leftarrow \Delta t / 2$;\quad reject and retry\;
    }{
        $(T_{\rm int},\, T_{\rm eff}) \leftarrow$ \AtmBC{$S',\, Y',\, Z',\, g$}\;
        $\Delta t \leftarrow \min\!\big(f_{\rm err}\, \Delta t,\; \Delta t_{\rm max}\big)$ \tcp*{grow timestep if error small}
        Save profiles; $(S, Y, Z, P, \rho, T, r_b) \leftarrow (S', Y', Z', P', \rho', T', r_b')$\;
        $t \leftarrow t + \Delta t$\;
    }
}
\caption{The \textsc{orchard} planetary evolution algorithm. Each step couples a Newton--Raphson transport solve for entropy $S$, helium $Y$, and metal $Z$ (\texttt{transport.py}) with a Henyey hydrostatic solver (\texttt{hydrostatic.py}). Adaptive error-controlled timestepping halves $\Delta t$ on failure and grows it when the maximum fractional change is well below tolerance.}
\label{alg:evolution}
\end{algorithm*}

\clearpage

\subsection{Setting Up Initial Models}\label{subsec:initial_setup}

This section describes how the initial ($t=0$) thermal and compositional structures of \orchard models are constructed and the degrees of freedom available to the user. The initialization is shared among \txt{initial.py} (mass mesh, initial entropy and composition profiles, and EOS instantiation), \txt{rk4\_initial.py} (the initial hydrostatic background), and the first, initialization call to the Henyey solver in \txt{hydrostatic.py} (thermal seeding of compact cores), corresponding to the blue boxes of Figure~\ref{fig:flowchart}. We first summarize the default workflow for a general model, then describe how inhomogeneous (``fuzzy-core'') initial states are specified, and conclude with the thermal seeding of compact rock/iron cores.

\subsubsection{Adiabatic/Homogeneous Gas Giant model set up}\label{subsubsec:general_init}

Generally, any gas giant model requires a desired planet total mass (\txt{M\_MJup} or \txt{M\_Mearth}). These are set to 1.0 M$_J$ by default in \txt{parameters\_default.ini}, and users can change this total mass directly. A general homogeneous and adiabatic model requires only a set of initial entropy (\txt{S\_ini}, in $k_B$ baryon$^{-1}$ units), helium mass fraction (\txt{Y\_ini}), and heavy element values (\txt{Z\_ini}). By default, the code will assume homogeneous structures at the given values. Once set, the model will be initialized at the specified \txt{S\_ini}, and cool adiabatically until the specified age is reached, set by the \txt{final\_age} parameter.

\subsubsection{Gas Giant Fuzzy core model set up.}\label{subsubsec:fuzzy_init}

Beyond homogeneous and adiabatic interior structures, users can specify structured initial profiles of $S$, $Y$, and $Z$, via the \txt{initial\_profiles\_S}, \txt{initial\_profiles\_Y}, and \txt{initial\_profiles\_Z} parameters.  To specify your own piecewise profile, select each of these to be \txt{struct}. Other functional forms are \txt{gaussian}, \txt{exponential}, and \txt{sigmoid} shapes constructed in \txt{utils/profiles.py}. The user provides the \txt{struct} profiles with coordinates along the $S$, $Y$, or $Z$ profile (e.g., \txt{z\_init\_coords}) as fractions of the total mass, ordered from the center (0) to the surface (1), together with the values at each breakpoint (\txt{z\_init\_values}). The provided break points are connected either linearly or with monotone PCHIP interpolation (\txt{struct\_interp}). As an example, the fuzzy-core initial state of a Jupiter-mass model (provided in the repository as \txt{parameter\_user\_jup\_inhomog\_struct\_example1.ini}) is specified as:

\begin{lstlisting}[language=bash]
[initial]
# Planet mass [Jupiter masses]
M_MJup = 1.0
# Outer (surface) values
S_ini = 9.0
Y_ini = 0.277
Z_ini = 0.017
# Build the initial structure via RK4 shooting
rk4_initial = True
# Breakpoints [center -> surface] in
# normalized mass coordinate
z_init_coords = [0, 0.2, 0.6, 1.0]
z_init_values = [0.15, 0.15, 0.017, 0.017]
s_init_coords = [0, 0.2, 0.6, 1.0]
s_init_values = [8.0, 8.0, 9.0, 9.0]
\end{lstlisting}

This configuration example sets the heavy element mass fraction at $Z = 0.15$ over the inner 20\% of the mass coordinate, grades it smoothly down to the surface value $Z = 0.017$ between $m/M_p = 0.2$ and $m/M_p = 0.6$, and retains $Z = 0.017$ over the outer 40\%. The entropy profile is treated analogously. The \txt{gaussian}, \txt{exponential}, and \txt{sigmoid} shapes provide analytic alternatives controlled by, e.g., \txt{init\_gauss\_center} and \txt{init\_gauss\_stdev\_z}. Every initialization parameter is documented individually in \txt{parameter\_descriptions.md} in the GitHub repository.

\subsubsection{Rocky interior model set up}\label{subsubsec:core_seeding}

The total mass is generally set by the \txt{M\_MJup} or \txt{M\_Mearth} parameters. We provide both since \txt{M\_Mearth} is more convenient for sub-Neptune, super-Earth, and terrestrial models. The ``core'' mass in gas giants, set by  \txt{mass\_core} can be set to be the same or a closer fraction of the \txt{M\_Mearth} parameter to produce either a sub-Neptune or super-Earth model. For example, setting \txt{M\_Mearth = 10} and \txt{mass\_core = 9.99} produces models with 1\% envelope/atmosphere mass fraction, akin to the thin envelope models presented in Figures~\ref{fig:supearths_0.5_10} and \ref{fig:3_5_melt_demo}. Setting \txt{mass\_core = 10} would produce a ``bare'' model with no atmosphere, cooling into space. The inner iron core mass is set by \txt{mass\_core\_fe}. Following this example, setting \txt{mass\_core\_fe = 3.33} would create an iron core mass of 3.33 Earth masses. Thus, if the set up is \txt{M\_Mearth = 10}, \txt{mass\_core = 9.99}, and \txt{mass\_core\_fe = 3.33}, then the user would create a sub-Neptune or super-Earth model of 10 Earth masses total, with 1\% H-He-Z envelope mass fraction and a mantle to iron core ratio of 2:1, reproducing a similar model to that presented on the bottom panel of Figure~\ref{fig:3_5_melt_demo}. Alternatively, setting \txt{isothermal\_compact\_core = True} initializes the entire compact core isothermally at the envelope-base temperature.

The mantles and cores are initialized adiabatically at a constant entropy value, akin to the models described in Section~\ref{subsubsec:general_init}. In the case of mantles, however, the user can specify whether to ensure temperature continuity with the envelope/atmosphere temperature by setting \txt{env\_s\_start = True}, or set their own mantle entropy with \txt{mantle\_entropy}. The mantle entropy parameter sets a manual desired entropy, usually between 0.5--0.7 $k_B$ baryon$^{-1}$, equivalent to between 8,000 K to 20,000 K initial temperatures. The iron core is initialized with temperature continuity with the mantle. However, the user can specify an entropy offset to initialize the iron core at higher temperatures, if desired, set by \txt{fe\_core\_offset} (in $k_B$ baryon$^{-1}$, default 0). A small positive offset of 0.03 was added to the models in Figures~\ref{fig:3_5_melt_demo} and \ref{fig:mearth_melt_demo}, thereby initializing their iron cores hotter than the mantle.

\clearpage

\section{Appendix: Explanation of the Partially Molten Convective Fluxes}
\label{sec:melt_flux_derivation}

We first derive the viscous mixing-length velocity of Equation~\ref{eq:v_misc}
from the parcel picture. A convecting parcel of size equal to the mixing
length $l$ acquires, over its mean displacement $l/2$, the temperature
contrast of Equation~\ref{eq:delta_T} and hence the fractional density
contrast $\delta\,\Delta T/T = \delta\,(\nabla-\nabla_{\rm ad})\,l/(2H_p)$,
corresponding to a buoyancy force per unit volume of
$\rho\,g\,\delta\,(\nabla-\nabla_{\rm ad})\,l/(2H_p)$. In the inviscid limit
this buoyancy accelerates the parcel, and the resulting work--energy balance
yields Equation~\ref{eq:v_mlt}, $v \propto (\nabla-\nabla_{\rm ad})^{1/2}$.
In the viscous limit appropriate to mantle convection, inertia is negligible, and the parcel instead moves at the terminal velocity at which buoyancy is
balanced by a creeping-flow viscous drag, which we close as a force per unit
volume $8\rho\,\nu\,v/l^{2}$, following the modified-MLT approach of
\cite{Sasaki1986a} and \seI:
\begin{equation}\label{eq:visc_balance}
    \rho\,g\,\delta\,(\nabla-\nabla_{\rm ad})\,\frac{l}{2H_p}
    = \frac{8\,\rho\,\nu\,v_{\rm visc}}{l^{2}}
    \quad\Longrightarrow\quad
    v_{\rm visc} = g\,\delta\,(\nabla-\nabla_{\rm ad})\,\frac{l^{3}}{16\,\nu\,H_p}\,,
\end{equation}
which is Equation~\ref{eq:v_misc}. The prefactor 16 is therefore the product
of the creeping-flow drag coefficient (8) and the factor 2 from evaluating
the buoyancy at the mean displacement $l/2$ (Equation~\ref{eq:delta_T}), and
the \emph{linear} dependence on $\nabla-\nabla_{\rm ad}$, in contrast to
the square-root dependence of the inviscid case. The parcel size is identified with the mixing length, and we set
$l = H_p$.

Equations~\ref{eq:fi_melt}--\ref{eq:fconv_melt} come from the same
MLT statement as Equations~\ref{eq:fconv_apple} and
\ref{eq:fconv_visc}, namely $\mathcal{F} = \rho\,v\,\Delta q$
(Equation~\ref{eq:f_conv_mlt}). Here, $v$ is the fluid parcel velocity associated
with its density contrast $\Theta \equiv \delta\rho/\rho$ over one
mixing length ($l = H_p$) and $\Delta q$ is its heat exchange, evaluated at the parcel mean displacement $l/2$ (Equation~\ref{eq:delta_T}). The inviscid/viscous limits differ only in the velocity law:
$v_{\rm invisc} = \sqrt{g\,\Theta\,H_p/8}$ (Equation~\ref{eq:v_mlt}) and
$v_{\rm visc} = g\,\Theta\,H_p^2/(16\,\nu)$ (Equation~\ref{eq:v_misc}).

A parcel in a partially molten zone can store an entropy perturbation in two
ways. In the inviscid or viscous channel, it is stored in
$\Delta s = (C_p/T)\,\Delta T$, the heat carried is $\Delta q = C_p\,\Delta T$,
and the buoyancy follows from thermal expansion,
$\Theta = \delta\,(\nabla - \nabla_{\rm ad})$. In the latent channel, the
parcel remains pinned to the melt curve at $T_{\rm melt}(P)$, so the
perturbation is stored in the melt fraction instead. The solidification entropy
jump gives $(\partial s/\partial \chi)_P = \mathcal{L}/T_{\rm melt}$, accounted for in Equation~\ref{eq:entropy_master}. Converting the entropy contrast accumulated over a full mixing length, $\Delta s \simeq -H_p\,(dS/dr)$, into a melt-fraction contrast gives $\Delta\chi = -(H_p\,T_{\rm melt}/\mathcal{L})\,(dS/dr)$. Analogous to the thermal component, the buoyancy entering the velocity laws uses this full contrast, $\Theta = \alpha_\chi\,\Delta\chi$, with $\alpha_\chi \equiv -(1/\rho)(\partial\rho/\partial\chi)_P$, while the heat exchanged is evaluated at the parcel mean displacement $l/2$ and therefore carries the factor $1/2$ of Equation~\ref{eq:delta_T}, $\Delta q = \tfrac{1}{2}\,\mathcal{L}\,\Delta\chi$. The inviscid flux is then
\begin{equation}\label{eq:melt_flux_assembled}
    \mathcal{F} = \rho\,v_{\rm invisc}\,\Delta q
    = \rho\,\sqrt{\frac{g\,\alpha_\chi\,\Delta\chi\,H_p}{8}}\;\frac{\mathcal{L}\,\Delta\chi}{2}
    = \rho\,\mathcal{L}\,\sqrt{\frac{\alpha_\chi\,g\,H_p}{32}}\,\Delta\chi^{3/2}\,,
\end{equation}
which is Equation~\ref{eq:fi_melt}. The conversion from Equations~\ref{eq:fconv_apple} and \ref{eq:fconv_visc} to Equations~\ref{eq:fi_melt} and \ref{eq:fv_melt} can then be directly substituted
\begin{equation}\label{eq:melt_substitution}
    \big(\,C_p T,\;\delta,\;\nabla - \nabla_{\rm ad}\,\big)
    \;\longrightarrow\;
    \big(\,\mathcal{L},\;\alpha_\chi,\;\Delta\chi\,\big)\,,
    \qquad
    \Delta\chi = -\frac{H_p\,T_{\rm melt}}{\mathcal{L}}\,\frac{dS}{dr}\,.
\end{equation}

\clearpage

\bibliography{references}{}
\bibliographystyle{aasjournalv7}
\end{document}